\documentstyle[11pt]{article}
\renewcommand{\theequation}{\arabic{section}.\arabic{equation}}

\begin{document}
\title{ \bf{ Quantum Electrodynamics 
in the Light-Front Weyl Gauge } }
\author{ 
{\it J. Przeszowski} \\
Institute of Fundamental Technological Research \\ 	
Polish Academy of Science\\
Swietokrzyska 21, 00-049 Warsaw, Poland \\ 
{} \\
{\it H.W.L. Naus}\thanks{Present Address: 
Institut f\"ur Theoretische Physik,
University of Hannover,
Appelstr. 2, D-30167 Hannover, Germany.} \\
Max-Planck-Institut f\"ur Kernphysik \\
\& Institut f\"ur Theoretische Physik,
University of Heidelberg\\
Heidelberg, Germany\\
and \\
{\it A.C. Kalloniatis}\thanks{Present Address:
Institut f\"ur Theoretische Physik, University of Erlangen-N\"urnberg,
Staudtstr.7, D-91058 Erlangen, Germany.}  \\   
Max-Planck-Institut f\"ur Kernphysik \\
Postfach 10 39 80, 
D-69029 Heidelberg, Germany  } 
\date{January 18, 1996}
\maketitle

\begin{abstract}
We examine QED(3+1) quantised in the `front form'  
with finite `volume' regularisation, namely in 
Discretised Light-Cone Quantisation. 
Instead of the light-cone or Coulomb gauges, we impose the light-front
Weyl gauge $A^-=0$. The Dirac method is used  
to arrive at the quantum commutation relations 
for the independent variables. We  apply 
`quantum mechanical gauge fixing' to implement Gau{\ss}' law, 
and derive the physical Hamiltonian in terms of unconstrained variables. 
As in the instant form, this Hamiltonian
is invariant under global residual gauge transformations,
namely displacements. 
On the light-cone the symmetry manifests itself quite differently. 
\end{abstract} 
PACS number(s): 11.10.Ef,11.10.Gh,12.20.Ds \\
Preprint: MPI-H-V43-1995

\newpage
\section{Introduction} 
In recent years there has been something of
a resurgence in Hamiltonian field theory particularly
with reference to strong interaction physics and
gauge theory. This resurgence 
is partly in order to complement the insight
provided by action oriented approaches and the
numerical results of lattice gauge theory,
partly due to the intuition and experience which the solution
to an `old-fashioned' Hamiltonian eigenvalue problem
can give, and partly because it
has remained somewhat underdeveloped as a formal method within the
evolution of quantum field theory. 
With reference to gauge theories, recent successful
applications have been made to QCD in two dimensions
\cite{Hamrefs} and to various theories
in higher dimensions \cite{Hamrefs2}.   
Amongst this (incomplete) list, the additional
feature of light-cone quantisation on a null-plane surface of 
$x^+ \equiv \frac{1}{\sqrt{2}}(x^0 + x^3) = 0$ is quite prominent.
The often touted `simplicity of the light-cone vacuum'
is one reason for the hope that the fusing of
Dirac's `front form' approach to Hamiltonian dynamics  
\cite{Dir49} with QCD may lead to sensible results
for hadron spectroscopy within a `first principles'
calculation. 

All the advantages of a light-cone Hamiltonian
approach to gauge theory -- as a physically insightful method to use --
would be lost, of course, in an inappropriate choice
of gauge. For various reasons the light-cone gauge $A^+=0$ has
been the traditional choice in this context. For one,
it is one of the class of noncovariant gauges for which
Faddeev-Popov ghost decoupling can be argued in an
unregulated formalism. It also allows for an operator
implementation of the Gau{\ss} constraint, so that in some
sense the Fock space approach directly gives access to
the `physical space'. However, the standard arguments
turn out to be invalid when one seeks to regularise
the theory, especially in the infrared. The method of
Discretised Light-Cone Quantisation achieves this by
taking space to be of finite volume and imposing
respectively periodic and antiperiodic boundary conditions
on bosons and fermions. One finds firstly, that
the light-cone gauge is no longer accessible due to
the `gauging away' of a zero Fourier mode 
that actually is gauge invariant \cite{FNP81}.    
The closest gauge choice is then the light-cone Coulomb gauge
$\partial_- A^+=0$. In addition, one finds that the
zero modes satisfy, in general, nontrivial
constraint equations \cite{FNP81}. Entangled with the problem
of solving these equations is a choice of gauge to
remove the rather large freedom untouched by the gauge choice
\cite{KaP93,KaP94}. This problem has been solved,
for example in perturbative QED(3+1) in \cite{KaR94}
by introducing additional, and complementary, gauge constraints
within the light-cone Coulomb gauge. Excluding the purely
global zero modes, those that are space independent, 
the Hamiltonian has been worked out to lowest order in
weak coupling.   
  
The difficulty with this approach is that it is
not evident how to extend the method to
non-Abelian gauge theory, where the problem of
finding complementary gauge conditions to
completely eliminate (even topologically trivial)
gauge freedom nonperturbatively
is expected to be quite difficult. In particular,
the task of isolating the `fundamental modular
domain' \cite{vBa92} in this framework is the great
challenge. So at this juncture, it is fair to
ask: is the light-cone Coulomb gauge the most
convenient initial gauge choice in order to
write down the physical light-cone Hamiltonian
in terms of unconstrained variables?
  
Returning to the `instant form', where quantisation
is achieved on a space-like surface at $x^0=0$,
the Weyl or temporal gauge $A^0=0$ is arguably a
natural choice. Quantisation is `canonical'
and the Hamiltonian is easy to write down, even
in QCD. Indeed, it has been available for some time 	 
\cite{Bjo79}. The outstanding problem here 
has been the implementation of Gau{\ss}'
law on physical states. Some insight has been
achieved through works such as \cite{Fey81}
or \cite{Fri83}.
However, recently a conceptually elegant way has
been developed called `quantum mechanical gauge
fixing' \cite{LNOT94}. Here, Gau{\ss}' law is
implemented by a series of unitary transformations
on the Hilbert space and operators, including
the Gau{\ss} law operator, until it becomes essentially
trivial to implement the constraint. The result is
a non-trivial expression for the Hamiltonian in 
unconstrained variables (as one would actually hope!) 
acting in the physical
space. This Hamiltonian has now been written down for QCD but
remains difficult to solve \cite{LNT94}. Nonetheless, at this stage
approximations which do not break local gauge
invariance are possible. Indeed, in several
simplified settings some physics results have been
obtained \cite{LST95}.

The goal of the present work is to 
derive the physical light-cone Hamiltonian
for quantum electrodynamics in the light-front
Weyl gauge. That we succeed in this task
shows that the light-cone approach can indeed
be married to the method of \cite{LNOT94} for
implementing Gau{\ss}' law. A comment is in order
here: already in \cite{LNOT94} the method is
applied to quantisation on {\it space-like}
surfaces approaching arbitrarily close to the
null-plane. In the present work we 
quantise directly on the null-plane surface.
We shall see that the form of the results 
will be quite different, although we hope to
recover essentially the same physics. We return
to this point in the discussion.
The eventual hope would be that the 
corresponding light-cone Hamiltonian for QCD,
including all zero modes, will be simpler to
deal with due to the `simple' vacuum aspect of
the light-cone. However, this is still a distant
hope given that QED(3+1) will turn out to look
more complicated in the present approach.   

Consistent with conservation of difficulty, we
encounter a problem from the outset: 
front form Hamiltonian dynamics is a constrained
system even in non-gauge theory, and therefore quantisation
is already non-canonical. 
Therefore there is no avoiding something like
the Dirac method \cite{Dir64}, or analogous methods,  
in order to obtain consistent (Dirac) brackets for
the independent fields thereby enabling the writing
down of a quantum theory.  
It is precisely here that, as in
the light-cone Coulomb gauge approach, one must
carefully unravel the various sectors related
to normal mode, proper zero modes and global zero
modes. Here we are using the decomposition of
Fourier space developed in \cite{KaP94}.
The resultant formalism becomes complicated for precisely
this reason. In particular one must invert rather
non-trivial operator equations analogous to those
encountered in the instant form \cite{Bjo79}.
In one case we can give an explicit nonperturbative definition
for such an inversion. However, this does not exhaust the
the constraints one confronts, and in other cases we rely
on a formally defined Green's function which, as yet, we
have not been able to explicitly give nonperturbatively.

Nonetheless, the
Dirac procedure can be performed, leading to disentangling
of independent fields, their commutators and the 
Hamiltonian. The method of \cite{LNOT94} is then a somewhat tedious but
conceptually straightforward procedure, which we perform
here in the `light-cone gauge representation', here  
the most natural and possibly only choice, 
unlike the instant form case
\cite{LNOT94}. Gau{\ss}' law is thus
implemented and the Hamiltonian computed. 

Despite the abstract form of the representation,
some comparisons with other works can be made. Firstly, if the global
zero modes are suppressed in our calculation
we obtain the same Hamiltonian
as in the light-cone Coulomb gauge \cite{KaP94,KaR94}
where the latter work omits these modes of total zero
momentum (or spatially constant fields). In fact, we give
the complete Hamiltonian including {\it all} modes. Our work
represents the first case where all zero mode sectors in
a (3+1) dimensional gauge theory have been comprehensively
included in a unified calculation within the framework of
light-cone quantisation.
Secondly, we are able to demonstrate the existence
of a residual displacement symmetry related to
large gauge transformations. In instant form QED it
has been suggested that the photon can be interpreted
as a Goldstone boson associated with spontaneous
breaking of precisely this symmetry \cite{LNOT94}. Here, matters
are more involved due to the quite different manifestation
of this symmetry on the light cone. 

The paper is organised as follows. In the next section
we explain the procedure by which
we arrive at a consistent formulation of the quantum
theory in the presence of constraints. In section three
we use unitary transformations to implement Gau{\ss}' law
and arrive at the Hamiltonian acting in the physical space.
In section four we discuss the displacement symmetry.
There is a discussion and a
brief statement of conclusions at the end.

\setcounter{equation}{0}

\section{Quantisation}
{\underline {Notation.}}
In the following, the formalism will have a tendency to
become rapidly complicated. We therefore take the pain
from the outset to establish a notation that will keep
formulae as simple and as transparent to the physics
as much as possible.

Our light-cone convention is that of Kogut and Soper
\cite{KoS70}: 
$x^{\pm} \equiv \frac{1}{\sqrt{2}} (x^0 \pm x^3)$
with $\partial^{\pm} = \partial_\mp$. The light-cone
{\it time} variable is $x^+$ and since we shall work on
the surface $x^+=0$, it shall be suppressed in the
following. The transverse directions $(x_1,x_2)$
are represented by $x_{\perp}$. The combined space components
$(x^-,x_{\perp})$ are represented as ${\vec x}$.
Space is taken to be a `hypertorus' with $-L < x^- < L$
and $-L_{\perp} < x_{\perp} < L_{\perp}$. Bosons will be assigned
periodic and fermions antiperiodic boundary conditions.
This coupled with light-cone parity means that fermions
will have no zero momentum component in any direction.

However, the gauge bosons will have zero modes,
and it is these we now disentangle. 
We follow the structure introduced
in \cite{KaP94}. First, we denote the full photon field
as $V_\mu({\vec x})$. This can be expanded in Fourier
modes. We distinguish the {\it simple} zero mode
\begin{equation}
\int_{-L}^{+L} \frac{dx^-}{2L} V_\mu(x^-,x_\perp) \ , 
\end {equation}
via which we build the {\it normal mode} gauge field
\begin{equation}
A_\mu ({\vec x}) \equiv V_\mu ({\vec x}) - 
\int_{-L}^{+L} \frac{dx^-}{2L} V_\mu(x^-,x_\perp)   
\;.
\end{equation}
These degrees of freedom are known to represent the
usual propagating photons in the light-cone representation,
and as such we reserve the symbol $A_\mu$ to denote
them. From the simple zero mode one can 
build the totally space-independent
modes or {\it global} zero modes
\begin{equation}
q_\mu = \int_{-L}^{L} \int_{-L_\perp}^{L_\perp}
\frac{dx^- d^2x_\perp}{8 L L_\perp^2} V_\mu ({\vec x})
\end {equation}
where in future we shall write $d^3x$ for $dx^- d^2x_{\perp}$
and suppress the limits of integration. 
Evidently, the $q_\mu$
are $0+1$ dimensional fields, namely quantum mechanical
variables: thus the notation $q$.   
Finally, the simple and global zero modes can be used to build modes
with {\it no} $x^-$-dependence but {\it no constant} part in $x_\perp$: 
\begin{equation}
a_\mu (x_\perp) = \int_{-L}^{+L} \frac{dx^-}{2L} V_\mu(x^-,x_\perp)
 - q_\mu
\;.
\end{equation}
These are called the {\it proper} zero modes.
The decomposition is now complete, and in the sequel
we shall refer to one of the normal $A_\mu$, proper zero mode
$a_\mu$ and global zero mode $q_\mu$ sectors. 

With the above organisation of sectors we must now define
the corresponding delta functions. 
We adopt the notation that the {\it periodic}  three-dimensional
delta function is represented as $\delta^{(3)}({\vec x} - {\vec y})$,
which includes the zero modes.  
For the {\it antiperiodic} delta function, a subscript `a' is
appended: $\delta_a^{(3)}({\vec x} - {\vec y})$. The explicit difference
between these two objects can be easily seen by expanding in discrete
Fourier modes.
Next we must distinguish the delta functions appropriate
for each mode sector for periodic functions. Thus in the
normal mode sector we must subtract the $x^-$ independent part
of $\delta^{(3)}$, and so define
\begin{equation}
\delta^{(3)}_n ({\vec x} - {\vec y}) \equiv 
\delta^{(3)}({\vec x} - {\vec y}) - 
{1 \over {2L}} \delta^{(2)} (x_\perp - y_\perp)
\; .
\end{equation}
In the proper zero mode sector, which lives
in the two-dimensional perpendicular space,
we must subtract the overall two-dimensional volume factor
in defining the relevant delta distribution:
\begin{equation}
\delta_p^{(2)}(x_\perp - y_\perp) \equiv \delta^{(2)}(x_\perp - y_\perp) - 
{1\over {4L_\perp^2}}
\; .
\end{equation}
One can show that these satisfy completeness in each of their
respective sectors.

Now one can safely invert differential operators  
such as $\partial_-$ and $\partial_\perp^2$ in terms of
well-defined Green's functions, taking care of the
respective mode sector in which the operator acts.  
It will be necessary to be fairly general here. With
the covariant derivative $D_\mu = \partial_\mu + i e V_\mu$, we define  
${\cal G}_{(-)}[{\vec x},{\vec y}; V_-]$ as the operator-valued
Green's function to $(i D_-)$:
\begin{equation}
(i D_-^x) {\cal G}_{(-)}[{\vec x},{\vec y}; V_-] \equiv
\delta^{(3)}({\vec x} - {\vec y}) - [Sub.]  
\;,
\end{equation}
where $[Sub.]$ denotes possible subtractions corresponding to zero
eigenvalues of the operator in question. For example, in case of the
periodic Green's function with zero functional argument, $[Sub]=
\frac{1}{2L}\delta^{(2)}(x_\perp - y_\perp)$. 
The perpendicular part of the Green's function is actually
just $\delta^{(2)}(x_\perp - y_\perp)$ and could be factored out,
but keeping it in the present form makes resulting expressions simpler
to write down. The operator aspect
of ${\cal G}_{(-)}$ can be understood, for example,
order by order in perturbation theory by building
with the Green's function to $(i\partial_-)$, namely
${\cal G}_{(-)}[{\vec x},{\vec y};0]$, which only
makes sense in the normal mode sector. 
A nonperturbative construction for this particular
Green's function is described in Appendix A.
The Green's function 
${\cal G}_{(\perp)}[x_\perp , y_\perp; {\cal O}]$ is defined by
the relation
\begin{equation}
[ \Delta_\perp^x - \frac{e^2}{2L} {\cal O}(x_\perp) ]
{\cal G}_{(\perp)}[x_\perp , y_\perp; {\cal O}]
\equiv  
 \delta^{(2)} (x_\perp - y_\perp ) 
\label{eq:Gperp}
\end{equation}
where $\Delta_\perp \equiv \partial_\perp^2$,
and now ${\cal O}$ is some field operator of mass dimension one.
Again, Eq.(\ref{eq:Gperp}) can be elucidated order by order in
perturbation theory for which one uses the
basic inversion of $\Delta_\perp$ via 
${\cal G}_{(\perp)}[x_\perp , y_\perp; 0]$. 
A nonperturbative definition is however nontrivial
and therefore the above operation is, at best, merely
formal and its concrete implementation remains an open problem.

These suffice to cover all the Green's functions we
will require in this work.
Later it will be useful to have a compact notation for
convolutions of the above Green's functions with other quantities
\begin{eqnarray}
\left({\cal G}_{(\perp)}[{\cal O}] \ast {\cal P} \right)(\vec{x}) 
& \equiv &
\int d^2y_\perp \ {\cal G}_{(\perp)}[x_\perp, y_\perp ; {\cal O}]\ 
{\cal P}(y_\perp, x^{-})\\
\left({\cal G}_{(-)}[{\cal O}] \ast {\cal P} \right)(\vec{x}) &\equiv&  
\int d^3{y} \ {\cal G}_{(-)}[\vec{x}, \vec{y}; {\cal O}]\ 
{\cal P}(\vec{y})   
\end{eqnarray}
for field operators ${\cal O}$ and ${\cal P}$. 

We now correspondingly decompose the canonical momenta
\begin{equation}  
\Pi^\mu_V({\vec x}) = \frac{\delta { L}}{\delta\left(\partial_+
V_\mu (\vec{x}) \right)} 
\end{equation}
into the three sectors of normal, proper zero and
global zero modes. We shall give the Lagrangian explicitly shortly.
We distinguish the following momenta
\begin{eqnarray}
\Pi^\mu & \equiv & \frac{\delta {L}}{\delta(\partial_+ A_\mu)} \\
\pi^\mu & \equiv & \frac{\delta {L}}{\delta(\partial_+ a_\mu)} \\
p^\mu & \equiv & \frac{\delta {L}}{\delta(\partial_+ q_\mu)}
\;.
\end{eqnarray}
Note that the integral expressions for $\pi$ and $p$ in terms 
 of $\Pi^\mu_V$ do {\it not} explicitly contain the factors
$1/(2L)$, $1/(8LL_\perp^2)$, respectively. 
Again, the notation in the global zero mode sector emphasises
its resemblence to quantum mechanics.

Turning to fermions, $\psi$, as usual we decompose in
spinor space using projectors $\Lambda^{\pm} \equiv \gamma^0
\gamma^{\pm} \ / \sqrt{2}$ 
leading to bispinors $\psi_\pm \equiv \Lambda^\pm \psi$.
Since the fermions are taken to be antiperiodic there is
no need to distinguish different Fourier sectors in these.
 
This completes the introduction to the basic structures 
needed in the formalism. Further definitions are given,
where needed, in the course of calculation or in the Appendices.

\noindent
{\underline {Reaching a Canonical Formulation.}}
The QED Lagrangian, expressed in terms of the complete fields
$V_\mu$, $\psi$ and $\psi^{\dagger}$, takes the standard form 
\begin{equation}
L = \int d^3{x} \left[ - \frac 1 4 (\partial_\mu V_\nu
- \partial_\nu V_\mu)(\partial^\mu V^\nu
- \partial^\nu V^\mu) + \bar{\psi}(i \gamma^{\mu} \partial_\mu -
e \gamma^\mu V_\mu - m)\psi\right] \;.\label{QEDLagr}
\end{equation}
Once the boson field is decomposed into its different sectors 
\begin{equation}
V_\mu(\vec{x}) = A_\mu (\vec{x}) + a_\mu(x_\perp) + q_\mu \;,
\end{equation}
the Lagrangian Eq.(\ref{QEDLagr}) breaks into three parts
\begin{equation}
L = L_{nm}  + L_{pzm} + L_{gzm} \;,\label{defLagrQED}
\end{equation}
where
\begin{eqnarray}
L_{nm} & = & \int d^3{x} \left[ - \frac 1 4 (\partial_\mu A_\nu
- \partial_\nu A_\mu)(\partial^\mu A^\nu
- \partial^\nu A^\mu) + \bar{\psi}(i \gamma^\mu \partial_\mu -
e \gamma^\mu A_\mu - m)\psi\right] \nonumber\\
&&\label{nmLagr}\\
L_{pzm} & = & \int d^3{x} \left[ - \frac 1 4 (\partial_\mu a_\nu
- \partial_\nu a_\mu)(\partial^\mu a^\nu
- \partial^\nu a^\mu) - e \bar{\psi} \gamma^\mu \psi a_\mu
\right]\label{pzmLagr}\\ 
L_{gzm} & = & (8LL_\perp^2) \frac 1 2 (\partial_{+}
q_{-})^2 - e q_\mu \int d^3{x} \bar{\psi} \gamma^\mu 
\psi \;.
\label{gzmLagr}
\end{eqnarray}
These formulae in turn can be simplified by decomposing
the total electromagnetic current, 
${\cal J}^\mu = - e {\bar \psi} \gamma^\mu \psi$, into its
normal mode and proper and global zero mode parts:
\begin{equation}
{\cal J}^\mu = J^\mu + j^\mu + Q^\mu
\; .
\end{equation}
This is given in detail in Appendix B.

Now we may outline our method of quantisation. The Weyl gauge
condition will be imposed strongly, namely we set $A_+ =
a_+ = q_+ = 0$ at the classical level. As usual, this
means Gau{\ss}' law (given explicitly later)
appears as a constraint to be
imposed on states. However, mixed in with this constraint
will be many other constraints peculiar to the light-cone
approach. In this section we will eliminate these (second
class) constraints while leaving the (first class)
Gau{\ss} constraint untouched. Our procedure for doing this
is as follows: We will analyse
different subsystems, where only one field sector is treated in
terms of independent degrees of freedom while the remaining fields 
are regarded as non-dynamical external fields and/or currents. Then
we will exchange non-dynamical modes for effective
interactions of dynamical ones and will give a simpler (though
non-local) Lagrangian where only dynamical fields are present.
This will result in a sequence of equivalent effective
Lagrangians which contain fewer modes but have 
Euler-Lagrange equations the same as those which are generated
by the primary Lagrangian provided the constraint equations are
implemented for nondynamical fields.
This procedure is based on the
observation that different Lagrangians can lead to the same
system of Euler-Lagrange equations though they may have very different
constraint structure (in a different context see, for example,
\cite{DeJ84}). 
One can feel free to choose the most
suitable one for carrying out the canonical quantization
procedure. 
In more detail, we shall start with the proper zero mode Lagrangian 
Eq.(\ref{pzmLagr}) and analyse its canonical structure. When the
Dirac brackets are found then another equivalent Lagrangian will be  
proposed which contains only dynamical modes (these having non-zero
brackets) and effective non-local terms. A similar analysis can
be done for the Lagrangian Eq.(\ref{nmLagr}) and another effective
Lagrangian for dynamical modes can be found. Then the sector of
global zero modes and fermions are analysed leading to
the final effective Lagrangian.  Gathering these
partial effective Lagrangians we obtain a total effective
Lagrangian which contains only the Gau{\ss} constraint. However,
the brackets/commutators will mostly take the noncanonical
form usually seen in the light-front. This is a consequence
of the implementation in this method of the typically light-front
second class constraints for the original Lagrangian.

\noindent
\underline{Step 1: Proper zero mode sector.}
The Lagrangian Eq.(\ref{pzmLagr}) can be written explicitly in  
light-front coordinates 
\begin{equation}
L^{pzm}_{Weyl} = \int d^3{x} \left[ 
- \partial_i a_{-} \partial_{+}a_i + \frac 1 2 (
\partial_{+}a_{-})^2 - \frac 
1 4 (\partial_{i}a_j - \partial_{j}a_i)^2 
+  j^\perp a_\perp + j^{-} a_{-} \right]\;,\label{defLagrpzm}
\end{equation}
where the Weyl gauge condition $a_+ = 0$ has been explicitly
imposed and the proper zero modes of currents are given 
explicitly in Appendix B.
This Lagrangian leads to the classical equations of motion
\begin{eqnarray}
\partial^2_{+} a_{-} & = & \partial_i 
\partial_{+} a_{i}   + j^{-}
\label{ELeq1}\\
- \partial_i \partial_{+} a_{-} & = &
\Delta_\perp a_i - \partial_i \partial_{k}
a_{k}  + j^i\label{ELeq2}
\end{eqnarray}
and the canonical conjugate momenta 
\begin{eqnarray}
\pi^{-}(x_\perp) & = &\displaystyle \frac{\delta
L^{pzm}_{Weyl}}{\delta (\partial_{+} a_{-}(x_\perp))} =
2 L \partial_{+} a_{-}(x_\perp)  \\ 
\pi^{i}(x_\perp) & = & \displaystyle \frac{\delta
L^{pzm}_{Weyl}}{\delta (\partial_{+} a_{i}(x_\perp))} = -
2 L \partial_{i}a_{-}(x_\perp)\label{defpi}  
\end{eqnarray}
for the independent gauge fields. Though our gauge choice has
eliminated one primary constraint connected with the gauge
potential $a_{+}$, there is still another constraint
Eq.(\ref{defpi}) and we must implement the Dirac 
constraint procedure here. The constraint Eq.(\ref{defpi}) appears in the
extended Hamiltonian density via Lagrange multiplier fields
$u_i(x_\perp)$ 
\begin{eqnarray}
{\cal H}_{E}^{pzm} & = & \frac{1}{2L} \partial_{+} a_{-} \pi^{-} 
+ \frac{1}{2L}  \partial_{+} a_i \pi^i - {\cal
L}^{pzm}_{Weyl} + u_i(\pi^i + 2L \partial_{i}a_{-}) \nonumber\\ 
& = &\frac 1 2 \left(\frac{\pi^{-}}{2L}\right)^2 + \frac 1 4 \left(
\partial_i a_j - \partial_j a_i\right)^2 - a_{-} j^{-} - a_i
j^i + u_i(\pi^i + 2 L \partial_{i}a_{-}) \;
\end{eqnarray}
and there is a sequence of constraints 
\begin{eqnarray}
\phi_1^i & = &  \pi^i + 2 L \partial_{i}a_{-} \simeq 0\label{phione}\\
\phi_2^i & = & \Delta_\perp a_i - \partial_i \partial_j a_j +
j^i + \frac{1}{2L} \partial_i \pi^{-}\simeq 0 \; ,\label{phitwo} 
\end{eqnarray}	
which allow one to eliminate dependent degrees of freedom. While
Eq.(\ref{phione}) can be easily solved for $\pi^i$, 
Eq.(\ref{phitwo}) is more involved. Its transverse projection
forms a constraint on the transverse part of $a_i$ which can
be trivially solved to give:
\begin{equation}
a^T_i = {\cal G}_{(\perp)}[0] \ast j^T_i
\label{transa}
\; .
\end{equation}
Here, the superscript $T$ indicates the transverse projection.
The longitudinal part of $a_i$ is {\it not}
constrained by Eq.(\ref{phitwo}).
This suggests it is part of a dynamical field. It is
convenient to define a new field, $\pi(x_\perp)$, 
in terms of $a^L_i$, the
longitudinal projection of $a_i$, via a decomposition
\begin{equation}
a_i^L = {\cal G}_{(\perp)}[0] \ast (
j^L_i - {1\over{2L}} \partial_i \pi)
\; .
\label{longa}
\end{equation}
This equation defines the field $\pi$.
One can thus reorganise the decomposition $a_i = a^T_i + a^L_i$
by substituting Eqs.(\ref{transa},\ref{longa}), and obtain
(writing the expression explicitly)
\begin{equation}
a_i(x_\perp) = - \int d^2y_\perp {\cal G}_{(\perp)}[x_\perp,
y_\perp; 0] \left (j^i(y_\perp) + \frac{1}{2L}\partial_i
\pi(y_\perp)\right)\;.  \label{transvproj}
\end{equation}
It will turn out later on that the field $\pi$ has no explicit
dependence on $j^i$ and is a good candidate for the canonical variable.
Next, the longitudinal projection of the constraint Eq.(\ref{phitwo}) 
can be solved for $\pi^{-}$
\begin{equation}
 \pi^{-}(x_\perp) = 2L
\int d^2y_\perp {\cal G}_{(\perp)}[x_\perp, y_\perp; 0]
\partial_j j^j(y_\perp) \;.\label{piminus}
\end{equation}
The Dirac brackets for independent fields are nonzero only for
the pair $(\pi, a_{-})$
\begin{equation}
\left \{\pi(x_\perp), a_{-}(y_\perp) \right \}_D
=  - \delta_p^{(2)} (x_\perp - y_\perp) 
\end{equation}
and the Dirac Hamiltonian at the end of
step one contains the effective nonlocal terms 
\begin{eqnarray}
H_D^{pzm}  & = &  \frac{2L}{2} \int d^2x_\perp \int d^2y_\perp j^i(x_\perp)
{\cal G}_{(\perp)}[x_\perp, y_\perp; 0] j^i(y_\perp) \\
& - & \ \int d^2x_\perp \int d^2y_\perp \pi(x_\perp) {\cal
G}_{(\perp)}[x_\perp, y_\perp; 0] \partial_i j^i (y_\perp) -
2L \ \int d^2x_\perp a_{-}(x_\perp) j^{-}(x_\perp) . \nonumber
\end{eqnarray}
One can check that the effective equations of motion which follow
from the above Dirac Hamiltonian and brackets
agree with the Euler-Lagrange equations Eqs.(\ref{ELeq1}-\ref{ELeq2}).
Therefore our Dirac Hamiltonian and brackets describe the same
classical system as the primary Lagrangian Eq.(\ref{defLagrpzm})
and one can give an equivalent Lagrangian density  
\begin{equation}
{\cal L}^{pzm}_{eff} = \frac{1}{2L} \pi \partial_{+} a_{-} -
{\cal H}_D^{pzm} = \frac{1}{2L} \pi \partial_{+} a_{-} +
\frac{1}{2L} \pi {\cal G}_{(\perp)}[0]\ast \partial_i j^i +
a_{-} j^{-} - \frac 1 2 j^i {\cal G}_{(\perp)}[0]\ast j^i \;,
\label{pzmefflagr} 
\end{equation}
which straightforwardly leads to the correct Dirac brackets and
equations of motion.

\noindent
\underline{Step 2: Normal mode sector.}
In the second step, let us analyse the sector of normal modes
of gauge field potentials ${A}_\mu$ treating normal modes
of electromagnetic currents $J^\mu$ as arbitrary external
sources. From the Lagrangian Eq.(\ref{nmLagr}) we take these terms
which contain ${A}_\mu$  
\begin{equation}
{\cal L}_{Weyl}^{nm} = \partial_{+} {A}_i \left(\partial_{-}
{A}_i - \partial_i A_{-}\right) 
+ \frac 1 2 \left(\partial_{+} {A}_{-} \right)^2 - \frac 1 4
\left( \partial_i {A}_j - 
\partial_j {A}_i\right)^2 + {A}_{-} {J}^{-} + {A}_i {J}^i \;,
\end{equation}
and find the Euler-Lagrange equations of motion 
\begin{eqnarray}
\partial_{+} \left( \partial_{+} {A}_{-}-
\partial_i {A}_{i}
\right) & = & {J}^{-} \label{eqmAminus}\\ 
(2 \partial_{+} \partial_{-} - \Delta_\perp) {A}_{i}
& = & \partial_{i} \left( \partial_{+} {A}_{-}
 - \partial_j {A}_{j} \right)  + {J}^{i} \label{eqmAi}
\end{eqnarray}
and the canonical momenta 
\begin{eqnarray}
\Pi^{-} & = &\displaystyle \frac{\delta
{L}_{Weyl}^{nm}}{\delta 
(\partial_{+} {A}_{-})} = \partial_{+}
{A}_{-} \\
{\Pi}^{i} & = & \displaystyle \frac{\delta
{L}_{Weyl}^{nm}}{\delta (\partial_{+} {A}_{i})} =
\partial_{-} {A}_{i} - \partial_{i}
{A}_{-}\;.  
\end{eqnarray}
Here there is only one constraint
\begin{eqnarray}
\Phi^i & = & {\Pi}^{i} - \partial_{-} {A}_i +
\partial_i {A}_{-} \simeq 0\label{Piicons}
\end{eqnarray}
which is second class, and the unconstrained variables have
non-trivial Dirac brackets
\begin{eqnarray}
\left \{{A}_{-}(\vec{x}),
{\Pi}^{-}(\vec{y}) \right \}_D & = & \delta_n^{(3)}
(\vec{x} - \vec{y})\label{Dbapi}\\ 
\left \{{A}_{i}(\vec{x}),
{\Pi}^{-}(\vec{y}) \right \}_D & = &
\frac i 2 \partial_i^x {\cal G}_{(-)}[\vec{x}, \vec{y}; 0]\\ 
\left \{{\Pi}^{-}(\vec{x}),
{\Pi}^{-}(\vec{y}) \right \}_D & = & 
 \frac i 2 \Delta_\perp {\cal G}_{(-)}[\vec{x}, \vec{y}; 0]\\
\left \{{A}_{i}(\vec{x}), {A}_j (\vec{y}) \right \}_D & = & -
\frac i 2 \delta_{ij} {\cal G}_{(-)}[\vec{x}, \vec{y}; 0]\;. 
\label{Dbaa} 
\end{eqnarray}
The Dirac Hamiltonian 
\begin{equation}
{\cal H}_D^{nm}{}' = \frac 1 2 ({\Pi}^{-})^2
+ \frac 1 4 \left( \partial_i {A}_j -
\partial_j {A}_i\right)^2 
- {A}_{-} {J}^{-}  - {A}_i {J}^i ,
\end{equation}
generates Hamilton's equations of motion for unconstrained
variables which agree with the Lagrange equations
Eqs.(\ref{eqmAminus}-\ref{eqmAi}). 
One may define a new field $\Pi = {\Pi}^{-} - \partial_i {A}_i$, 
which simplifies the structure of Dirac brackets 
\begin{eqnarray}
\left \{{A}_{-}(\vec{x}),
\Pi(\vec{y}) \right \}_D & = & \delta_n^{(3)}
(\vec{x} - \vec{y})\label{Dbapi1}\\ 
\left \{{A}_{i}(\vec{x}), {A}_j (\vec{y})
\right \}_D & = & -  \frac i 2 \delta_{ij} {\cal G}_{(-)}[\vec{x},
\vec{y}; 0] \;,
\end{eqnarray}
and then all other brackets vanish. This new Dirac Hamiltonian
\begin{equation}
{\cal H}_D^{nm} = \frac 1 2 (\Pi)^2 + \Pi \partial_i {A}_i + \frac 1
2 \left( \partial_i {A}_j \right)^2 - {A}_{-} {J}^{-} - {A}_i
{J}^i 
\end{equation}
generates equations of motion which are equivalent to the
previous ones and
can be used for defining an effective Lagrangian 
\begin{eqnarray}
{\cal L}^{nm}_{eff} & = & \partial_{+} {A}_{i}\partial_{-}
{A}_{i} + \Pi \partial_{+} {A}_{-} - {\cal H}_{D}\nonumber \\
& = & \partial_{+} {A}_{i}\partial_{-} {A}_{i} - \frac 1 2
\left( \partial_i {A}_j\right)^2 - \frac 1 2 (\Pi)^2 + \Pi
\left( \partial_{+} {A}_{-} - \partial_i {A}_i \right) + {A}_{-}
{J}^{-} + {A}_i {J}^i . \ \
\label{nmefflagr}
\end{eqnarray}
Having analysed the canonical structure of the gauge field sector
one can 
substitute the Lagrangian Eq.(\ref{pzmLagr}) by
Eq.(\ref{pzmefflagr}) and the boson part of Eq.(\ref{nmLagr}) by
Eq.(\ref{nmefflagr}) and instead of the total Lagrangian
Eq.(\ref{defLagrQED})  one can work with the effective Lagrangian  
\begin{eqnarray}
{\cal L}^\prime_{eff} & = & 
\partial_{+} {A}_{i}\partial_{-} {A}_{i} -
\frac 1 2 \left( \partial_i {A}_j\right)^2 - \frac 1 2 (\Pi)^2 +
\Pi \left( \partial_{+} {A}_{-} - \partial_{i} {A}_{i} \right)
+ \frac{1}{2L}  \pi \partial_{+} {a}_{-} 
\nonumber \\ 
& + & \frac 1 2 \left( \partial_{+}
q_{-}\right)^2  
 +  i \sqrt{2} {\psi_{+}}^{\dagger} \partial_{+} \psi_{+} + i
\sqrt{2} {\psi_{-}}^{\dagger} \partial_{-} \psi_{-} + i
{\psi_{-}}^{\dagger} \alpha^i \partial_{i} \psi_{+} + i
{\psi_{+}}^{\dagger} \alpha^i \partial_{i} \psi_{-}\nonumber\\
&- & m {\psi_{+}}^{\dagger} \gamma^0 \psi_{-} - m
{\psi_{-}}^{\dagger} \gamma^0 \psi_{+}-  e \sqrt{2}
{\psi_{-}}^{\dagger} \psi_{-}{V}_{-} \label{efflagr1}
\label{prelimeffLag} \\
&-& eV^{\prime}_i \left({\psi_{+}}^{\dagger} \alpha^i \psi_{-}+
{\psi_{-}}^{\dagger} \alpha^i \psi_{+}\right)- \frac 1 2 j^i
\left( {\cal G}_{(\perp)}[0] \ast j^i\right) \; . \nonumber 
\end{eqnarray}
In this expression, 
\begin{equation}
V^\prime_i = 
{A}_i - \frac{1}{2L}  \partial_i \left({\cal G}_{(\perp)}[0] \ast \pi
\right) + q_{i},
\end{equation}
namely it is the original $V_i$ but with 
its proper zero mode $a_i$ expressed as in
Eq.(\ref{transvproj}) and the new current-current interaction subtracted.
This piece has been explicitly reintroduced as the last
term in Eq.(\ref{prelimeffLag}).  The decomposition of $V_-$
remains unchanged.

\noindent
\underline{Step 3: Fermion sector.}
In the next step, we take the fermion part of the effective
Lagrangian Eq.(\ref{efflagr1}) 
\begin{eqnarray}
{\cal L}^{fer}_{Weyl} & = & i \sqrt{2} {\psi_{+}}^{\dagger}
\partial_{+} 
\psi_{+} + i \sqrt{2} {\psi_{-}}^{\dagger} \partial_{-} \psi_{-} +
i {\psi_{-}}^{\dagger} \alpha^i \partial_{i} \psi_{+} + i
{\psi_{+}}^{\dagger} \alpha^i \partial_{i} \psi_{-}\nonumber\\
&- & m {\psi_{+}}^{\dagger} \gamma^0 \psi_{-} - m
{\psi_{-}}^{\dagger} \gamma^0 \psi_{+}-  e \sqrt{2}
{\psi_{-}}^{\dagger} \psi_{-}{V}_{-} \label{fermionlagr}\\
& - &e{V}_i^{\prime}
\left({\psi_{+}}^{\dagger} \alpha^i \psi_{-}+ 
{\psi_{-}}^{\dagger} \alpha^i \psi_{+}\right)- \frac 1 2 j^i
\left( {\cal G}_{(\perp)}[0] \ast j^i\right) \;.\nonumber 
\end{eqnarray}
Now the system contains many more constraints and
these are given in Appendix C.
The nondynamical modes are the fermion field $\psi_{-}$
and the global zero mode $q_{i}$.
These are determined by the following differential equations and global
integral condition  
\begin{eqnarray}
&& \left( i \sqrt{2} \partial_{-} - e \sqrt{2} V_{-} 
\right) \psi_{-} = - i \alpha^i \partial_i
\psi_{+} + m \gamma^0 \psi_{+} \nonumber\\
&&\ \ + \ e \alpha^i \left( V_i - {\cal
G}_{(\perp)}[0] \ast {j^{i}} \right)
\psi_{+} \ , \label{eqpsim}\\ 
&& 0 = Q^i = \frac{1}{8LL^2_\perp} \int \ d^3 {x}\ (
\psi^{\dagger}_{+} \alpha^i \psi_{-} + \psi^{\dagger}_{-} \alpha^i
\psi_{+})(\vec{x}). 
\end{eqnarray}
First one gets
\begin{eqnarray}
\psi_{-}(\vec{x}) & = & \frac{1}{\sqrt{2}} \left({\cal
G}_{(-)}[V_{-}]\ast \xi\right)(\vec{x}) \nonumber\\
&& - \frac{e}{\sqrt{2}}  \left({\cal
G}_{(\perp)}[0]\ast j^i (x_\perp) - q_{i} \right) \alpha^i \left(
{\cal G}_{(-)}[V_{-}]\ast \psi_{+}\right)(\vec{x}) \;, \label{prsolpsim} 
\end{eqnarray}
where
\begin{eqnarray}
\xi(\vec{x}) & = & \left[ m \gamma^0 - i\alpha^i \partial_i +
e \alpha^i \left({A}_i(\vec{x}) - \frac{1}{2L} \partial_i
 \left( {\cal G}_{(\perp)}[0] \ast \pi\right)(\vec{x})  \right)\right]
\psi_{+}(\vec{x})\label{defxi} \; . 
\end{eqnarray}
Note that these are not yet solutions for the dependent fermion
field $\psi_{-}$ because these fields appear
also on the right hand side in the zero mode currents $j^i$.
However one can introduce them into the definition of $j^i +
Q^i$ (see Eqs.(\ref{defqi},\ref{defji})) 
\begin{equation}
j^i(x_\perp) + Q^i
= - \frac{e}{2 L} \Gamma^i(x_\perp) + \frac{e^2}{2L}
{\cal M}^{ik}(x_\perp) \left(({\cal G}_{(\perp)}[0]\ast
j^k)(x_\perp) - q_{k}\right) \;,
\end{equation}
where
\begin{eqnarray}
\Gamma^i(x_\perp) & = & \frac{1}{\sqrt{2}} \int dx^{-}
 \psi_{+}^{\dagger}(\vec{x}) \alpha^i \left({\cal G}_{(-)}[V_{-}] \ast
 \xi \right) (\vec{x}) \nonumber\\ 
& + &  \frac{1}{\sqrt{2}} \int dx^{-}
 \xi^{\dagger}(\vec{x}) \alpha^i \left({\cal G}_{(-)}[V_{-}] \ast
 \psi_{+} \right) (\vec{x})\\
{\cal M}^{ij}(x_\perp) & = & \frac{1}{\sqrt{2}} \int dx^{-}
 \psi_{+}^{\dagger}(\vec{x}) \alpha^i \alpha^j \left({\cal G}_{(-)}[V_{-}] \ast
 \psi_{+} \right) (\vec{x}) \nonumber\\ 
& + &  \frac{1}{\sqrt{2}} \int dx^{-}
 \psi^{\dagger}_{+} (\vec{x}) \alpha^j \alpha^i \left({\cal G}_{(-)}[V_{-}] \ast
 \psi_{+} \right) (\vec{x})\nonumber\\
&= & \delta^{ij} \sqrt{2} \int dx^{-}
 \psi^{\dagger}_{+} (\vec{x}) \left({\cal G}_{(-)}[V_{-}] \ast
 \psi_{+} \right) (\vec{x}) = \delta^{ij} {\cal M}^2(x_\perp) 
\; .
\label{gamM}  
\end{eqnarray}
Now, from the constraint $Q^i = 0$ one gets the differential equation 
\begin{equation}
\left[ \Delta_{\perp} - \frac{e^2}{2L} {\cal M}^2(x_\perp) \right]
\left(({\cal G}_{(\perp)}[0]\ast j^i)(x_\perp) - q_{i}\right) =
- \frac{e}{2L} \Gamma^i (x_\perp), \label{formalQisol}
\end{equation}
which has a formal solution 
\begin{equation}
({\cal G}_{(\perp)}[0]\ast j^i)(x_\perp) - q_{i} = 
 - \frac{e}{2L} \int
d^2y_\perp {\cal G}_{(\perp)}[x_\perp, y_\perp; {\cal M}^2]
\Gamma^i(y_\perp)
\end{equation}
in terms of the functional Green's function 
introduced in Eq.(\ref{eq:Gperp}).
Finally, we may express the nondynamical fermion field as
\begin{equation}
\psi_{-}(\vec{x})  = \frac{1}{\sqrt{2}} \left({\cal
G}_{(-)}[V_{-}]\ast \xi\right)(\vec{x}) + \frac{e^2}{2 \sqrt{2}
L} \left( {\cal G}_{(\perp)}[{\cal M}^2] \ast \Gamma^i\right)
(x_\perp) \left( {\cal G}_{(-)}[V_{-}]\ast
\psi_{+}\right)(\vec{x}) \;, \label{prsolpsi} 
\end{equation}
whereby we obtain 
\begin{equation}
H^{fer}_D  =  \frac{1}{\sqrt{2}} \int d^3x 
\xi^{\dagger}(\vec{x}) \left({\cal G}_{(-)}[V_{-}] \ast
\xi\right) (\vec{x}) + \frac 1 2 \frac{e^2}{2L} \int d^2x_\perp
\Gamma^i (x_\perp) \left( {\cal G}_{(\perp)} [{\cal M}^2] \ast
\Gamma^i \right) (x_\perp)\label{Dhamglob} 
\end{equation}
as the Dirac Hamiltonian for unconstrained fields. 
Just as before we can give the effective Lagrangian for the
fermions 
\begin{eqnarray}
{\cal L}_{eff}^{fer} & = & (\partial_{+} \psi_{+})
\Pi_{\psi_{+}} - {\cal H}^{fer}_D = i
\sqrt{2}\psi^{\dagger}_{+} \partial_{+} \psi_{+} \nonumber\\
& - & \frac{1}{\sqrt{2}} \xi^{\dagger} ({\cal G}_{(-)}[V_{-}]
\ast \xi ) -\frac 1 2 \frac{e^2}{4L^2} \Gamma^i 
\left({\cal G}_{(\perp)} [{\cal M}^2] \ast \Gamma^i \right)
\label{efflagrfer}. 
\end{eqnarray}

\noindent
\underline{Quantum theory.}
Having eliminated all nondynamical fields one may now 
proceed by substituting 
the fermion part of Eq.(\ref{efflagr1}) by that given
in Eq.(\ref{efflagrfer}). One thereby
obtains the total effective Lagrangian 
\begin{eqnarray}
{\cal L}_{eff} & = & \partial_{+}
{A}_{i}\partial_{-} {A}_{i} - 
\frac 1 2 \left( \partial_i {A}_j\right)^2 - \frac 1 2 (\Pi)^2 +
\Pi \left( \partial_{+} {A}_{-} - \partial_{i} {A}_{i} \right)
- \frac{1}{2L} \pi  \partial_{+} {a}_{-} + 
\frac 1 2 \left( \partial_{+}
q_{-}\right)^2 \nonumber\\ 
& +& i \sqrt{2}\psi^{\dagger}_{+} \partial_i \psi_{+} - \frac{1}{\sqrt{2}}
 \xi^{\dagger} ({\cal G}_{(-)}[V_{-}] \ast \xi )
-\frac 1 2 \frac{e^2}{4L^2} \Gamma^i \left({\cal
G}_{(\perp)} [{\cal M}^2] \ast \Gamma^i \right) 
\label{effQEDLagr} 
\end{eqnarray}
as the starting point for the canonical
quantisation procedure. It generates Euler-Lagrange equations
of motion which agree with the dynamical equations of the
primary Lagrangian Eq.(\ref{QEDLagr}) with all nondynamical equations
(formally) implemented. The canonical momenta and Dirac brackets are
the same as in the former partial analyses and are given in the
Appendix C. They lead to the equal-$x^{+}$ quantum commutation
relations 
\begin{eqnarray}
\left [{\Pi}(\vec{x}),
{A}_{-}(\vec{y}) \right ] & = & - i \delta_n^{(3)}
(\vec{x} - \vec{y})\label{comPiA-}\\ 
\left \{ \psi_{+}^{\dagger}(\vec{x}),
{\psi_{+}}(\vec{y}) \right \} & = & 
\frac{1}{\sqrt{2}} \ \Lambda^{(+)} \delta^{(3)}_a(\vec{x}- \vec{y})\\
\left [\pi(x_\perp), {a}_{-} (y_\perp) \right ] & = & - i
\delta_p^{(2)} (x_\perp - y_\perp)\label{comaLa-}\\ 
\left [{A}_{i}(\vec{x}), {A}_j (\vec{y}) \right ] & = &
  \frac{\delta_{ij}}{2} {\cal G}_{(-)}[\vec{x},\vec{y};0]\\
\left[ p^{-}, {q}_{-} \right ] & = & - i
\end{eqnarray}
and the quantum Hamiltonian, like ${\cal L}_{eff}$, which comes from 
${\cal H}_D^{pzm}$, ${\cal H}_D^{nm}$ and $ H_D^{fer}$, 
\begin{eqnarray}
H_{eff} & =& \int d^3{x}\left[ \frac 1 2
(\Pi(\vec{x}))^2 + \Pi \partial_i {A}_i (\vec{x}) +
\frac 1 2 \left( \partial_i {A}_j(\vec{x}) \right)^2
 \right] \nonumber\\ 
& + &\frac{1}{16 L L^2_\perp} (p^{-})^2 + \frac{1}{\sqrt{2}}
\int d^3{x} \xi^{\dagger}(\vec{x}) \left({\cal G}_{(-)}[V_{-}]
\ast \xi\right) (\vec{x})
\nonumber \\
& + &\frac 1 2 \frac{e^2}{2L} \int d^2x_\perp \Gamma^i (x_\perp)
\left( {\cal G}_{(\perp)} [{\cal M}^2] \ast \Gamma^i \right)
(x_\perp), \label{quanHam}
\end{eqnarray} 
where the formal expressions must be defined by some careful
ordering procedure. Due to the effective equations of motion 
the Gau{\ss}' law operator 
\begin{equation}
G(\vec{x}) = \partial_{-} \Pi(\vec{x}) + 2 \partial_{-} \partial_i
{A}_i (\vec{x}) - \Delta_\perp {A}_{-}(\vec{x}) -
\Delta_\perp {a}_{-} (x_\perp) - e \sqrt{2}
\psi^{\dagger}_{+}(\vec{x}) \psi_{+}(\vec{x}) 
\label{gaussop}
\end{equation}
is
$x^{+}$-independent. It leads to a classically first class constraint,
$G \simeq 0$; namely it must annihilate physical states in the
quantum theory. Furthermore, it is intimately connected to the
residual gauge symmetry (see Appendix D).

\noindent
\underline{Translation Generators.}
With the effective Lagrangian density Eq.(\ref{effQEDLagr}) one
can calculate the canonical momentum-energy tensor
\begin{eqnarray}
T^{\nu\mu} & = & \frac{\delta {L}_{eff}}{\delta \ (\partial_{\nu} {A}_i)}
\partial^{\mu} {A}_i + \partial^{\mu} \psi_{+}
\frac{\delta {L}_{eff}}{\delta \
(\partial_{\nu} \psi_{+})} + \frac{1}{2L} \frac{\delta {L}_{eff}}{\delta \
(\partial_{\nu} \pi)} \partial^{\mu} \pi 
\nonumber\\ 
& + & \frac{1}{2L} \frac{\delta {L}_{eff}}{\delta \
(\partial_{\nu} a_{-})} \partial^{\mu} a_{-} + \frac{1}{8 L
L_\perp^2} \frac{\delta {L}_{eff}}{\delta \ (\partial_{\nu}
q_{-})} \partial^{\mu} q_{-} - g^{\nu \mu} {\cal L}_{eff}.
\end{eqnarray}
Then from the generators of translations $P^\mu = \int d^3{x}
\ T^{+\mu}(\vec{x})$, the spatial translations are 
\begin{eqnarray}
P^{i} &= &\int d^3{x} \left[ - \partial_{-}{A}_k
\partial_{i} {A}_k - {\Pi} \partial_{i}
{A}_{-} - i \sqrt{2} \psi^{\dagger}_{+} \partial_{i}
\psi_{+} - \frac{1}{2L}\pi \partial_i {a}_{-}
\right] \label{Momi}\\ 
P^{+} & = & \int d^3{x} \left[ \partial_{-} {A}_k
\partial_{-} {A}_k + {\Pi} \partial_{-}
{A}_{-} + i \sqrt{2} \psi^{\dagger}_{+} \partial_{-}
\psi_{+}\right]. \label{Momplus}
\end{eqnarray}
Now from the (anti)commutation relations
Eqs.(\ref{comPiA-}-\ref{comaLa-}) one can recover 
the correct Heisenberg relations for
all dynamical quantum fields $\varphi_J = ({A}_k,
{A}_{-}, {\Pi}, \pi, {a}_{-})$
\begin{eqnarray}
\partial_{i} \varphi_J = - i \left[ P^{i},\varphi_J
\right]\\ 
\partial_{-} \varphi_J = i \left[ P^{+},\varphi_J
\right]
\end{eqnarray}
and this confirms the translation invariance of QED in the Weyl
gauge.
We note that the generators $P^{+}$ and $P^{i}$ are not invariant
under the residual gauge transformation with 
gauge function $h({\vec x})$ 
\begin{eqnarray}
P^{+}_h & = & \Omega_{h} P^{+} \Omega^{\dagger}_{h} = P^{+}
+ \int d^3{x} \ G(\vec{x}) \partial_{-} h(\vec{x})  \\
P^{i}_h & = & \Omega_h P^{i} \Omega^{\dagger}_h = P^{i}
- \int d^3{x} \ G(\vec{x}) \partial_{i} h(\vec{x}) 
\label{Motrafo}
\end{eqnarray}
and this is connected with the lack of gauge invariance of the
canonical energy-momentum tensor. 
We return to this below.

\setcounter{equation}{0}

\section{Implementing Gau{\ss}' law} 
In the previous section we derived the canonical formulation
of light-front QED. The quantum Hamiltonian Eq.(\ref{quanHam})
is supplemented with Gau{\ss} law operator Eq.(\ref{gaussop}), 
which is to be implemented as a constraint on the
physical states,
\begin{equation}
G(\vec{x})  |{\rm{phys}}\rangle = 0 .
\label{gauss} 
\end{equation}
Being the generator of residual gauge transformations (see Appendix D)
the Gau{\ss} law operator commutes with the Hamiltonian and, consequently, time
evolution leaves the physical space invariant. The presence of 
Eq.(\ref{gauss}) means that there are still redundant variables
in the theory, which, in principle, can be eliminated in the
physical sector of Hilbert space. 
We now apply `quantum mechanical gauge fixing' \cite{LNOT94}
to the light-front formulation in order to arrive at the 
`light-cone gauge representation' (to be defined below) of
the physical Hamiltonian.

The general principle of this method is to construct a unitary gauge
fixing transformation, which acts as a gauge transformation
on the degrees of freedom to be kept in the theory. The gauge function, 
however, is a functional of the field variable to be eliminated from
the Hamiltonian.  The appropriate choice of the functional
indeed will make  the transformed Hamiltonian independent of
that variable, i.e. it becomes 'cyclic'. Concomitantly, due
to the underlying gauge invariance, the transformed
Gau{\ss} law operator enables one to eliminate the
corresponding conjugate momentum in the physical space.
It turns out that the most convenient variable to
treat in this manner is the $V_-$ field. For this reason, the
final result will be said to be in `light-cone gauge representation'.

\noindent
\underline{First unitary gauge fixing transformation.}
We first choose the normal mode $A_-$ as the variable to be
eliminated and so define the following unitary gauge fixing 
transformation
\begin{equation}
U_1[\vartheta] = 
\exp\left( - i \int d^3 x g({\vec x}) \vartheta[{\vec x}; A_-]\right)  , 
\label{firsttrans}
\end{equation}
with
\begin{equation}
g({\vec x}) \equiv G({\vec x}) - \partial_- \Pi({\vec x})
= 2 \partial_- \partial_i A_i - \Delta_\perp A_- - 
 \Delta_\perp a_-  - e \sqrt{2} \psi^\dagger_+ \psi_+   
\end{equation}
and 
\begin{equation}
\vartheta [{\vec x};A_-] = i \int d^3y {\cal G}_{(-)}[\vec{x},
\vec{y}; 0] A_-(\vec{y}) .
\end{equation}
Obviously, $U$ acts as a gauge transformation on the transverse
gauge fields,
\begin{equation}
U_1 A_j U_1^\dagger = A_j + \partial_j \vartheta  ,
\end{equation}
zero mode gauge fields,
\begin{equation}
U_1 a_- U_1^\dagger = a_-  ,
\end{equation}
\begin{equation}
U_1 \pi U_1^\dagger = \pi - \Delta_\perp
\int_{-L}^{L} dy^{-}\ \vartheta[y^{-}, x_\perp;A_-] = \pi  ,
\end{equation}
 and the fermions
\begin{equation}
U_1 \psi_+ U_1^\dagger = \exp(- i e \vartheta ) \psi_+ .  
\end{equation}
It leaves $A_-$ invariant. Due to the $A_-$ dependence
of the gauge function the transformation of $\Pi$ is
non-trivial 
\begin{equation}
U_1 \Pi U_1^\dagger = \Pi - i ({\cal G}_{(-)}[0]\ast g)({\vec x}) - 
2 \Delta_\perp \vartheta  .
\end{equation}
The coefficient of $\Delta_\perp \vartheta$
 in the last term stems from the non-commutativity
of $g(\vec{x})$ and $g(\vec{y})$.
The transformation of the operators appearing in the fermionic
part of the Hamiltonian effectively eliminates $A_-$ from them
\begin{equation}
U_1 \ \int d^3{x} \
\xi^{\dagger}(\vec{x}) \left({\cal G}_{(-)}[V_-] \ast
\xi\right) (\vec{x}) \ U_1^\dagger =
\int d^3{x} \ \xi^{\dagger}(\vec{x}) \left({\cal
G}_{(-)}[a_{-}+q_-] \ast \xi\right) (\vec{x}) . 
\end{equation}
The transformed Gau{\ss} law operator reads
\begin{equation}
U_1 G U_1^\dagger = \partial_- \Pi + {1\over{2L}} \int dx^- g({\vec x}). 
\end{equation}
By integrating over $x^-$, one can project out
two separate constraints on the transformed physical states
\begin{equation}
\Pi |{\rm{phys}}^\prime\rangle = 0 
\label{newgauss1} 
\end{equation}
and
\begin{equation}
\int { {dx^-}\over{2L}} g({\vec x})  |{\rm{phys}}^\prime\rangle = 0, 
\label{newgauss2} 
\end{equation}
where
\begin{equation}
 |{\rm{phys}}^\prime \rangle \equiv  U_1  |{\rm{phys}}\rangle .
\end{equation}
The first constraint Eq.(\ref{newgauss1})
can readily be implemented at this point. 
Now we have all the ingredients to calculate the
transformed Hamilton operator acting in the physical sector
of Hilbert space. The result, after the first
unitary gauge fixing transformation, is 
\begin{equation}
U_1 H U_1^\dagger |{\rm{phys}}^\prime\rangle =  
H^\prime |{\rm{phys}}^\prime\rangle ,  
\end{equation}
with 
\begin{eqnarray}
H^\prime & =& \int d^3{x}\left[ \frac 1 2
\left( \partial_j A_j - i \sqrt{2}e {\cal G}_{-}[0] \ast
(\psi^\dagger_+\psi_+ )
\right)^2 +
\frac 1 4 \left( \partial_i A_j -\partial_j A_i\right)^2
 \right] \nonumber\\ 
& + &\frac{1}{16 L L^2_\perp} (p^{-})^2 + \frac{1}{\sqrt{2}} \int
d^3{x} \xi^{\dagger}(\vec{x}) ({\cal G}_{(-)}[a_{-}+q_-] \ast
\xi)(\vec{x}) \nonumber \\
& + &\frac 1 2 \frac{e^2}{2L} \int d^2x_\perp \Gamma^{\prime i} (x_\perp)
\left( {\cal G}_{(\perp)} [{\cal M^\prime}^2] \ast \Gamma^{\prime i} \right)
(x_\perp) .
\label{Hprime}
\end{eqnarray} 
The operators $\Gamma^\prime$ and ${\cal M^\prime}$ can be obtained
from the operators $\Gamma$ and ${\cal M}$ by replacing $V_-$  with
$a_- + q_-$ in their respective definitions, namely Eq.(\ref{gamM}).
Because of the fact that it is impossible to invert completely
the operator $\partial_-$  a residual Gau{\ss} law operator 
makes its appearance through Eq.(\ref{newgauss2}), 
namely we have the operator  
\begin{equation}
G_2(x_\perp)  = -  \Delta_\perp a_-(x_\perp) - 
          {1\over{2L}} \rho_2 (x_\perp) , 
\end{equation}
where the two dimensional charge density is defined by
\begin{equation}
 \rho_2= \sqrt{2}e \int dx^- \psi^\dagger_+ \psi_+.
\end{equation}
$G_2$ commutes with the Hamiltonian and generates $x^{+}$ {\it and}
$x^{-}$ independent gauge transformations.  Explicitly one has
\begin{eqnarray}
\Omega_2[\gamma] &=& \exp\left(i\int d^2x_\perp 2L G_2(x_\perp)
\gamma(x_\perp) \right) \nonumber \\
\mbox{} &=& \exp\left(-i \int d^3x  \left( \Delta_\perp a_- (x_\perp)
  + \sqrt{2}e \psi^\dagger_+ (\vec{x})
\psi_+(\vec{x}) \right) \gamma(x_\perp)\right) ,
\label{Gtra2}
\end{eqnarray} 
which yields
\begin{equation}
\Omega_2 \psi_+ (\vec{x}) \Omega_2^\dagger  = \exp(ie\gamma(x_\perp))
\psi_+(\vec{x}),
\end{equation}
and 
\begin{equation}
\Omega_2 \pi(x_\perp) \Omega_2^\dagger = \pi(x_\perp) + {2L}
\Delta_\perp \gamma (x_\perp). 
\end{equation}
The other operators are invariant, and the invariance of $H$ is
easily checked. 
Let us also consider the momentum operators 
(given by Eqs.(\ref{Momi}-\ref{Momplus})).
Straightforward but rather lengthy calculations yield in the
physical sector of Hilbert space
\begin{equation}
U_1 P^i U_1^\dagger = \int d^3 x \left( -\partial_- A_j \partial_i A_j
-i\sqrt{2} \psi^\dagger_+ \partial_i \psi_+ - 
\frac{1}{2L}\pi \partial_i a_- \right),
\end{equation}
and
\begin{equation}
U_1 P^+ U_1^\dagger = \int d^3 x \left( \partial_- A_j \partial_- A_j
+i\sqrt{2} \psi^\dagger_+ \partial_- \psi_+ \right) .
\end{equation}
The latter is invariant under the residual gauge tranformations
given by Eq.(\ref{Gtra2}),
\begin{equation}
\Omega_2 P^+ \Omega_2^\dagger = P^+,
\end{equation}
but for the transverse components one has
\begin{equation}
\Omega_2 P^i \Omega_2^\dagger = P^i - 2L\int d^2x_\perp
G_2(x_\perp) \partial_i \gamma(x_\perp).
\end{equation}

It is instructive to recall the derivation in \cite{LNOT94}:
there the `axial gauge representation' was given for the 
equal-time formulation of QED. Up to this point it is rather analogous 
with the present case if 
one identifies the `$-$' components with the `$3$' components of gauge and
electric fields. 

\noindent
\underline{Second unitary gauge fixing transformation.}
The condition Eq.(\ref{newgauss2}) reflects that there remain
redundant variables in the theory.  A second unitary gauge fixing 
transformation is in order and the strategy remains in principle the same.  
However, now a striking
difference between the light-cone and equal-time formulation appears.  
In the equal time framework the residual Gau{\ss} law is independent
of the zero modes fields in the $3$--direction, which means that these
degrees of freedom cannot be eliminated, i.e. are physical. In contrast,
on the light--cone the residual Gau{\ss} law still contains
$a_-$; in fact (apart from the fermionic charge density) it {\it only}
contains that particular zero mode. Excluding unitary gauges involving
fermions, this leaves no other choice as the complete light--cone 
gauge representation.  This means we will eliminate $a_-$
and its conjugate momentum $\pi$ via a second unitary
gauge fixing transformation and resolution of the residual
Gau{\ss} law operator.

Applying the same principle as above, we define the transformation
\begin{equation}
U_2= \exp\left (i\int d^2x_\perp \rho_2(x_\perp) \eta[x_\perp;
\pi]\right),
\end{equation}
where $\eta$ simply is 
\begin{equation}
\eta[x_\perp; \pi] = \frac{-1}{2L} ({\cal G}_{(\perp)}[0]
\ast\pi)(x_\perp) . 
\end{equation}
It acts as a gauge transformation in the fermionic sector
\begin{equation}
U_2 \psi_+ U_2^\dagger = \exp(-ie\eta) \psi_+.
\end{equation}
It leaves $\pi$ invariant and transforms $a_-$ as follows
\begin{equation}
U_2 a_-(x_\perp) U_2^\dagger = a_-(x_\perp) - 
\frac{1}{2L}  \left(
{\cal G}_{(\perp)}[0] \ast 
\rho_2 \right)(x_\perp).
\label{atra2}
\end{equation}
In the fermionic sector  we obtain 
\begin{equation}
U_2 \xi(\vec{x}) U_2^\dagger = \exp\left(-ie\eta[x_\perp; \pi]\right) 
\chi(\vec{x}) = \exp\left(i \frac{e}{2L} ({\cal
G}_{(\perp)}[0] \ast\pi)(x_\perp)  \right) \chi(\vec{x}) , 
\end{equation}
with
\begin{equation}
\chi = [ m\gamma_0-i\alpha^i\partial_i +e \alpha^i A_i] \psi_+.
\end{equation}
Now an additional complication arises
in the transformation of the appearing composite operators. 
For instance,
upon transforming the first $\xi$-dependent term in 
Eq.(\ref{Hprime}) 
one explicitly encounters the expression 
\begin{eqnarray}
U_2  \int d^3 x
\xi^{\dagger}(\vec{x}) {\cal G}_{(-)}[a_{-}+q_-] \ast
 \xi(\vec{x}) \
U_2^\dagger = 
\hspace{3.8cm}  
 \nonumber\\
 \int d^3 x
\chi^{\dagger}(\vec{x}) \exp\left(ie \eta[x_\perp; \pi]\right) 
\left({\cal G}_{(-)}[a_{-}- \frac{1}{2L} 
{\cal G}_{(\perp)}[0] \ast \rho_2 \ + \ q_-]
\ast \exp(-ie\eta[x_\perp; \pi]) \chi\right)(\vec{x})
\; .
\nonumber \\
\end{eqnarray}

Similar expressions hold for 
the operators $\Gamma^\prime$ and ${\cal M^\prime}$.
Since $a_{-}$ and $\pi$ do not commute, we cannot cancel
the exponential functions in order to eliminate $\pi$
without an additional contribution: the zero mode field
$a_{-}$ gets shifted by a singular object, i.e.
\begin{equation}
\exp\left(ie \eta[x_\perp; \pi]\right) a_-(x_\perp)
\exp\left(-ie\eta[x_\perp; \pi]\right)=a_-(x_\perp) 
-\frac{e}{2L}{\cal G}_{(\perp)}[x_\perp, x_\perp ;0].
\end{equation}
Here we implicitly assume some regularisation of the
small distance singularity involved in
${\cal G}_{(\perp)}[x_\perp,x_\perp;0]$. 
It is, however, not necessary to be
more explicit since this singular {\it c-number} appears everywhere in
combination with the global zero mode $q_-$. Hence it is possible
to absorb the singular term into a redefinition of this global
zero mode: 
\begin{equation}
q_-' = q_- - \frac{e}{2L}{\cal G}_{(\perp)}[x_\perp, x_\perp ;0].
\end{equation}
Obviously the new global zero mode, $q_-'$, has the same
commutation relations 
as $q_-$. This procedure completely cures the potential problem in
eliminating $\pi$ from the Hamiltonian. 
Eq.(\ref{atra2})
immediately yields for the residual Gau{\ss} law operator
\begin{equation}
U_2 G_2 U_2^\dagger = - {\cal G}_{(\perp)}[0] \ast a_- 
+\frac{\sqrt{2}e}{2L} \Delta_\perp 
 \int dx^- {\cal G}_{(\perp)}[0] \ast 
\left( \psi^\dagger_+ \psi_+ \right) 
-\frac{1}{2L} \rho_2 
\end{equation}
Again it separates into two constraints on the transformed physical
states
\begin{equation}
a_-(x_\perp)  |{\rm{phys''}} \rangle =0,
\label{aminusconstr}
\end{equation}
and
\begin{equation}
Q |{\rm{phys''}} \rangle = \int d^2 x \rho_2(x_\perp)  
 |{\rm{phys}''} \rangle =0,
\label{globQ}
\end{equation}
with
\begin{equation}
|{\rm{phys''}} \rangle \equiv U_2 |{\rm{phys}}^\prime \rangle .
\end{equation}
Eq.(\ref{globQ}), the neutrality
condition, will be the only constraint left in this
formulation (out of the infinitely
many we had at the start).
The other constraint, Eq.(\ref{aminusconstr}), can be readily 
implemented at this point. 
Note that the transformed
Hamiltonian does not depend on the conjugate variable $\pi$
anymore. In this way we finally arrive at the following Hamiltonian
\begin{eqnarray}
H_{{fin}} & =& \int d^3x\left[ \frac 1 2
\left( \partial_j A_j - i \sqrt{2} e
{\cal G}_{(-)}[0] \ast ( \psi^\dagger_+\psi_+)
\right)^2 +
\frac 1 4 \left( \partial_i A_j -\partial_j A_i\right)^2
 \right] \nonumber\\ 
& + &\frac{1}{16L L^2_\perp} (p^{-})^2 + \frac{1}{\sqrt{2}} \int
d^3{x} \chi^{\dagger}(\vec{x}) \left({\cal G}_{(-)}\left[ q^\prime_- -
{\cal G}_{(\perp)}[0] \ast \rho_2\right] \ast \chi\right) (\vec{x}) \nonumber \\
& + &\frac 1 2 \frac{e^2}{2L} 
\int d^2x_\perp \Gamma^{\prime \prime i} (x_\perp)
\left( {\cal G}_{(\perp)} [{\cal M}^{\prime \prime 2}] \ast
\Gamma^{\prime \prime i} \right)
(x_\perp)  \;,
\label{finalHamil}
\end{eqnarray} 
acting in the physical sector of the Hilbert space.
The operator $\Gamma^{\prime \prime i}$ is defined in the same way
as $\Gamma$ (cf. Eq.(\ref{gamM})), however
with $\xi$ and $V_-$ replaced by 
$\chi, q^\prime_- - {\cal G}_{(\perp)}[0] \ast \rho_2$  
respectively. 
The second replacement also yields the operator
${\cal M}^{\prime \prime}$
starting from the definition of ${\cal M}$.  Apart from
the global constraint, Eq.(\ref{globQ}), which eventually projects on the
charge zero sector, all the constraints are implemented. In other
words, $H_{{fin}}$ is indeed formulated in terms of unconstrained
degrees of freedom. Eq.(\ref{finalHamil}) is the main result
of this paper. 

\noindent
{\underline {The Translation Generators.}}
Finally, we return to the momentum operators.
Although they were not gauge invariant in the large Hilbert space it
is clear from Eqs.(\ref{Motrafo},\ref{gauss}) that they are invariant
in the physical sector.
Therefore one can already anticipate that, finally,
they can be expressed in terms of physical variables only.
Indeed, we obtain in the physical space
\begin{equation}
U_2 U_1 P^i U_1^\dagger U_2^\dagger =
- \int d^3 x \left( \partial_- A_j \partial_i A_j
+i\sqrt{2} \psi^\dagger_+ \partial_i \psi_+ \right) 
\equiv P^i_{{fin}} ,
\end{equation}
and
\begin{equation}
U_2 U_1 P^+ U_1^\dagger U_2^\dagger
= \int d^3 x \left( \partial_- A_j \partial_- A_j
+i\sqrt{2} \psi^\dagger_+ \partial_- \psi_+ \right)  
\equiv P^+_{{fin}},
\end{equation}
as the generators of translations.

\setcounter{equation}{0}

\section{Displacement Symmetry}

Apart from the gauge transformations considered so far, the canonical
Weyl gauge Hamiltonian, cf. Eq.(\ref{quanHam}), is invariant
under displacements. 
These are gauge transformations {\it not} generated by Gau{\ss}' law. 
This displacement symmetry is given by  the following unitary operator
\begin{equation} 
\Omega_d = \exp \left[- i \int d^3{x}  \left( e \sqrt{2} \psi^{\dagger}(\vec{x})
\psi(\vec{x}) + p^- \partial_{-} \right) \vec{\beta}
\cdot \vec{x}\right] \;,
\label{disp}
\end{equation} 
where 
\begin{equation}
\beta_{-} = \frac{\pi}{e\ L} n, \ \ \ n = \pm 1, \pm 2, \ldots
\ \ \beta^{i} = \frac{\pi}{e \ L_\perp} m^i \ \ \ m^i = \pm 1,
\pm 2,\ldots \ . \label{betas}
\end{equation} 
Under displacements only the fermion fields and the zero mode
gauge  field variable, $q_-$, transform
\begin{eqnarray}
\Omega_d \psi_{+}(\vec{x}) \Omega_d^{\dagger} & = & e^{i\ e\ \vec{\beta}
\cdot \vec{x}}  \psi_{+}(\vec{x})\label{psi+displ}\\
\Omega_d q_- \Omega_d^{\dagger} & = &
q_{-} - \beta_{-}.\label{qminusdispl}
\end{eqnarray} 
Compatibility of Eqs.(\ref{psi+displ}-\ref{qminusdispl}) with the boundary
conditions for fermion and boson fields, respectively, forces
the form of the gauge function $\beta^\mu$ given in Eq.(\ref{betas}).
Now the transformation properties of composite fermion operators
follow immediately
\begin{eqnarray}
\Omega_d \xi_{+}(\vec{x}) \Omega_d^{\dagger} & = & e^{i\ e \ \vec{\beta}
\cdot \vec{x}}\left(  \xi_{+}(\vec{x}) + e\alpha^i \beta^i \psi_{+}(\vec{x})
\right) \\
\Omega_d \Gamma^i(x_\perp) \Omega_d^{\dagger} & = & \Gamma^i(x_\perp) +
e \beta^i {\cal M}^2(x_\perp).
\end{eqnarray} 
We can now see how the 
Hamiltonian of Eq.(\ref{quanHam}) changes under displacements.
Only its fermionic ($\xi$ dependent) part transforms 
in a non--trivial way:
\begin{equation}
\Omega_d H^{eff}_{fer} \Omega_d^{\dagger}  = H^{eff}_{fer}
+ \delta_1 H^{eff}_{fer} + \delta_2 H^{eff}_{fer} \;,
\end{equation} 
where $\delta_1 H^{eff}_{fer}$ and $\delta_2 H^{eff}_{fer}$
are linear and quadratic in $\vec{\beta}$ respectively
\begin{eqnarray}
\delta_1 H^{eff}_{fer} & = &  \frac{e}{\sqrt{2}} \int d^3{x}
\xi^{\dagger} \alpha^i \beta^i ({\cal G}_{(-)}[V_{-}] \ast
\psi_{+}) +  \frac{e}{\sqrt{2}} \int d^3{x}
\psi^{\dagger}_{+} \beta^i \alpha^i ({\cal G}_{(-)}[V_{-}] \ast
\xi) \nonumber\\
& + & \frac 1 2 \frac{e^3}{2L}\ \int d^2x_\perp
\beta^i \left[ {\cal M}^2 ({\cal G}_{(\perp)} [{\cal M}^2] \ast
\Gamma^i ) + \Gamma^i( {\cal G}_{(\perp)} [{\cal M}^2] \ast {\cal
M}^2) \right] \nonumber\\
& = & e \beta^i \int d^2x_\perp \left[
\Gamma^i + \frac{e^2}{2 L} {\cal M}^2 ({\cal G}_{(\perp)} [{\cal
M}^2] \ast \Gamma^i )\right] \label{delH1}\\
\delta_2 H^{eff}_{fer} & = &  \frac{e^2}{\sqrt{2}} \int d^3{x}
\psi^{\dagger}_{+}\alpha^j \beta^j \alpha^i \beta^i ( {\cal
G}_{(-)}[V_{-}] \ast \psi_{+} ) \nonumber \\
& + &
 \frac 1 2 \frac{e^4}{2L}\ \int d^2x_\perp
\beta^i \beta^i {\cal M}^2  ( {\cal G}_{(\perp)} [{\cal
M}^2] \ast {\cal M}^2) \nonumber\\
& = &\frac{e^2}{2} \beta^i \beta^i \int d^2x_\perp {\cal M}^2 
\left[ 1 + \frac{e^2}{2 L} ( {\cal G}_{(\perp)} [{\cal M}^2] \ast {\cal
M}^2)\right].
\label{delH2}
\end{eqnarray}
Using Eq.(\ref{formalQisol}) and periodic boundary conditions,
one can show that these vanish separately:
\begin{eqnarray}
\delta_1 H^{eff}_{fer} & = & e \beta^i \int d^2x_\perp 
\left[ \Gamma ^i + \frac{e^2}{2 L} {\cal M}^2 ( {\cal
G}_{(\perp)} [{\cal M}^2] \ast \Gamma^i )\right]\nonumber\\
& = &  e \beta^i \int d^2x_\perp  \Delta_\perp ( {\cal
G}_{(\perp)} [{\cal M}^2] \ast \Gamma^i ) = 0 \label{delH11}\\
\delta_2 H^{eff}_{fer} & = & \frac{e^2}{2} \beta^i \beta^i \int d^2x_\perp 
\left[ {\cal M}^2  + \frac{e^2}{2 L} {\cal M}^2 ( {\cal
G}_{(\perp)} [{\cal M}^2] \ast {\cal M}^2)\right]\nonumber\\
& = &  \frac{e^2}{2} \beta^i \beta^i \int d^2x_\perp  \Delta_\perp ( {\cal
G}_{(\perp)} [{\cal M}^2] \ast {\cal M}^2 ) = 0
\; .
\label{delH22}
\end{eqnarray} 
In this way we have explicitly established the invariance of $H$ under
displacements.

Since the displacements are not generated by Gau{\ss}' law one expects
this symmetry to be present also in the final Hamiltonian, namely
after quantum mechanically gauge fixing. First, one easily 
proves that the displacement operator is invariant under the gauge fixing
transformations
\begin{equation}
U_2 U_1 \Omega_d U_1^\dagger U_2^\dagger = \Omega_d.
\end{equation}
Secondly, whereas the infinitely many symmetries  generated by
Gau{\ss}' law essentially disappear because Gau{\ss}' law is implemented,
the displacement symmetry survives this process. In other words,
the local gauge transformations reduce to the identity while
the global displacement symmetry remains a non-trivial symmetry of 
$H_{{fin}}$.
With the analogous transformation properties as  given in 
Eqs.(\ref{delH1}-\ref{delH22}), 
one indeed verifies this invariance, i.e
\begin{equation}
\Omega_d H_{{fin}} \Omega_d^\dagger = H_{{fin}},
\end{equation}
as above. Furthermore, the momentum operators transform as follows
\begin{eqnarray}
\Omega_d P^+_{{fin}} \Omega_d^\dagger & = & 
P^+_{{fin}} -\beta_- Q, \\ 
\Omega_d P^i_{{fin}} \Omega_d^\dagger & = & P^i_{{fin}} +\beta^i Q.
\end{eqnarray}
Obviously, their shift is proportional to the total charge.
Therefore, the remaining global constraint, Eq.(\ref{globQ}), 
guarantees the compatibility of translation invariance and these
global residual gauge transformations in the 
physical space.

\setcounter{equation}{0}

\section{Discussion and Conclusions}
First let us summarise what we have achieved in the present work:
working in the light-front formalism for quantum electrodynamics,
we implemented the Weyl gauge $A^-=0$ and quantised the theory.
By putting the system in a finite `volume' the (possibly singular) infrared
behaviour was {\it a priori} regularised and all the
degrees of freedom could be cleanly disentangled into normal,
proper zero and global zero mode sectors. By performing unitary
transformations we succeeded in implementing Gauss law
and deriving the light-front Hamiltonian which acts in the physical
Hilbert space -- albeit in a somewhat abstract form. 
Again, we say abstract because we have not explicitly described the
nonperturbative meaning of the functional Green's function
${\cal G}_{(\perp)}$. 
We have simply assumed 
existence and uniqueness of the Green's function. Nevertheless,
the final result for the Hamiltonian is actually new, 
and has not been given, even at this `formal' level,
including the global zero modes. It is the inclusion of
these fields that allowed us to verify the invariance of
the theory under global (large gauge transformation) displacements.
We can thus confirm that the symmetry works just as well
on the light-front as it does in standard quantisation.

We now compare the significance of the displacement
symmetry in the present light-front formulation with
what was learned in \cite{LNOT94} for equal-time. 
On the light-cone only the `$-$' component of the photon field is
affected by the displacements. However, in the instant-form all
three components of the gauge field are shifted.  As was pointed out in 
\cite{LNOT94} the displacement symmetry actually can be understood from
Maxwell's equations by identifying the zero-mode of the
displacement vector as conserved quantity. In light-front coordinates
the relevant Maxwell equations read in full 
\begin{eqnarray}
\partial_{+} \left( \partial_{+} {V}_{-}-
\partial_i {V}_{i}
\right) & = & {\cal J}^{-} \ ,\\ 
(2 \partial_{+} \partial_{-} - \Delta_\perp) {V}_{i}
& = & \partial_{i} \left( \partial_{+} {V}_{-}
 - \partial_j {V}_{j} \right)  + {\cal J}^{i}  \ .
\end{eqnarray}
By means of the continuity equation we rewrite these as
\begin{eqnarray}
\partial_{+} \left( \partial_{+} {V}_{-}-
\partial_i {V}_{i}
\right) & = & \partial_-(x^-{\cal J}^{-})+\partial_i(x^-{\cal J}^i)
 +x^-\partial_+{\cal J}^+ \ , \\ 
(2 \partial_{+} \partial_{-} - \Delta_\perp) {V}_{i}
& = & \partial_{i} \left( \partial_{+} {V}_{-}
 - \partial_i {V}_{i} \right)  + \partial_-(x^i {\cal J}^-)+
\partial_j(x^i {\cal J}^j) + x^i\partial_+ {\cal J}^+ .
\end{eqnarray}
Integration over space and dropping surface terms
such as $x^- {\cal J}^-|_{-\infty}^{\infty}$, namely
assuming the current to be localised, leads to
the two equations
\begin{eqnarray}
\partial_+^2 q_- & = & \int d^3x \ x^- \partial_+ {\cal J}^+ \ ,\\
0  & = & \int d^3x \ x^i \partial_+ {\cal J}^+. 
\end{eqnarray} 
(Of course, one can be more precise concerning the omission of surface
terms and impose periodic boundary conditions \cite{LNOT94}. This yields
the conservation law in exponentiated form, corresponding to 
Eq.(\ref{disp}),
but for our present purpose this difference is not relevant.)
In this way we obtain the following conserved quantities
\begin{eqnarray}
\partial_+ \left( p^- -  \int d^3x \ x^- {\cal J}^+ \right) &=& 0 \ , \\
\partial_+ \int d^3x \ x^i {\cal J}^+ &=& 0. 
\end{eqnarray} 
Indeed only the light-cone component of the electric field,
namely $p^-$, appears in the conserved quantities since the transverse
zero mode components are not dynamical. In particular, this means that
for free Maxwell theory there is only {\it one}
non-trivial displacement symmetry.

The latter feature obscures an interpretation analogous to the one in the
equal-time theory. There it was argued that, for zero or weak coupling,
the displacement symmetry is spontaneously broken,
giving rise to zero-mass particles:
the physical photons. The argument for free Maxwell theory was based
on the formal analogy with free massless scalar theory, where
the massless scalars can indeed be interpreted as Goldstone bosons
\cite{ItZ85}. 
Actually, even for free massless scalar light-cone theory
the interpretation of the masslessness in terms of the Goldstone
mechanism is unclear.  This may indicate either a possible pathology
in massless non-interacting theories on the light-cone or
the need for further studies of the Goldstone mechanism and its
consequences in that framework. We note one work in this direction
by \cite{KTY95}.  

Of course, in the interacting light-cone theory the situation may  
actually be similar to equal-time  QED. In other words, the 
displacement symmetries may be spontaneously  broken leading
to photons as Goldstone particles. This is supported by
the empirical fact that photons are massless.
However, to see this in, say, perturbation theory is difficult
given the above problems already at {\it zeroth} order.
Up to now, other theoretical arguments are also lacking.
Nevertheless, the formal developments presented here, can serve
as the starting point for such interesting investigations.

\section*{Acknowledgements}
One of us (JP) thanks MPI for hospitality and DFG-PAC Exchange
Programme for financial support 
while visiting Heidelberg where part of this
work was completed. (ACK) and (HWLN) were supported by 
a {\it Max-Planck Gesellschaft} Stipendium. (ACK) also
thanks the Department of Physics of the University of Tasmania
for its hospitality while this work was completed.

\setcounter{section}{0}

\setcounter{equation}{0}

\renewcommand{\theequation}{\Alph{section}.\arabic{equation}}
\renewcommand{\thesection}{}
\section{Appendix A. Construction of Green's Function ${\cal G}_{(-)}$.  }
To construct the Green's function ${\cal G}_{(-)}$
we follow the procedure outlined in \cite{LNT94}
and give the eigenfunctions to the operator
$i D_-$:
\begin{eqnarray}
 iD_- \zeta_n(\vec{x}) & = & \lambda_n (x_\perp) \zeta_n(\vec{x}),
\label{opereigeneq} \\
\zeta_n(\vec{x}) & = & 
\exp\left[-ie\left(\tau(\vec{x})-x^- v(x_{\perp}) +\frac{2\pi nx^-}{2Le}
+\Delta(x_{\perp})\right)\right], \label{eigenfun} \\
\lambda_n (x_\perp) & = & \frac{2\pi n}{2L} -ev(x_{\perp}) 
\label{eigenval}\\
\tau(\vec{x}) & = & \int_{-L}^{x^-} dz^- \; V_-(z^-,x_{\perp}), \\
v(x_{\perp}) & = & \frac{1}{2L} \int_{-L}^{L} dx^- \; V_-(\vec{x}), 
\end{eqnarray}
with $\Delta$ an arbitrary local phase and $n$ any integer.
A useful relation to show that $\zeta_n$ and $\lambda_n$
satisfy Eq.(\ref{opereigeneq}) is
\begin{equation}
i\partial_- \zeta_n(\vec{x}) = e\left[ V(\vec{x})-v(x_{\perp})
+ \frac{2\pi n}{2eL}\right] \zeta_n(\vec{x})
\end{equation}
which itself follows easily from Eq.(\ref{eigenfun}).
We see that both the eigenfunctions $\zeta_n$ and eigenvalues
$\lambda_n$ are operator valued. The theory is Abelian
and $[V_-({\vec x}), V_-({\vec y})] = 0$ so there are no ordering
ambiguities. The phase $\Delta$ can be chosen such that the
$V_-$ dependence in $\zeta_n$ is given by just the periodic
step function $\tilde{\vartheta}$, used in \cite{LNOT94},
namely
\begin{equation}
\Delta(x_{\perp})= \int_{-L}^{L} dy^- \; 
\frac{y^- -L}{2L} V_-(y^-,x_{\perp}) .
\end{equation}

The eigenfunctions are, by construction, periodic.
As one can explicitly verify, this set of eigenfunctions is 
orthonormal and complete
\begin{eqnarray}
\frac{1}{2L} \int_{-L}^{L} dx^- \; \zeta_n^* (\vec{x}) \zeta_m (\vec{x})
& = & \delta_{nm} \;, \\
\frac{1}{2L} \sum_n \zeta_n(x^-,x_{\perp}) \zeta_n^*(y^-,x_{\perp})
& = & \delta(x^- - y^-) \;.
\end{eqnarray}

Let us now consider the differential equation
\begin{equation}
iD_- F = K \;
\end{equation}
where $F$ and $K$ are periodic functions. They can thus be
expanded in the eigenfunctions $\zeta_n$. Then we readily obtain
a relation for the respective
expansion coefficents $f_n(x_{\perp}), k_n(x_{\perp})$
of $F$ and $K$:
\begin{equation}
\lambda_n f_n = k_n \;.
\label{eigenvalues}
\end{equation}
For non-zero eigenvalues this equation is trivially solved. If 
eigenvalue zero appears in the spectrum, $\lambda_{n_{0}} = 0$, 
we conclude the following: First, the zero mode 
(with respect to the covariant derivative)
of $K$, i.e. $k_{n_0}$, must vanish in order that the differential
equation indeed has solutions. Secondly, even in that case the zero mode
of $F$, $f_{n_{0}}$, cannot be solved for from Eq.(\ref{eigenvalues}).
Assuming that, either there is no eigenvalue zero, or that if
$\lambda_{n_0} = 0$ then $k_{n_0} = 0$, leads to the explicit
solution
\begin{equation}
F(\vec{x}) = \sum_{n \neq n_0}  \frac{\zeta_n(\vec{x})}{2L\lambda_n}
\int_{-L}^{L} dy^- \; \zeta_{n}^{\dagger}(y^-,x_{\perp}) 
K(y^-,x_{\perp}) .
\end{equation}
Thus we can identify a Green's function
\begin{equation}
{\cal G} (x^- , y^-; x_\perp) = \sum_{n \neq n_0} 
\frac{\zeta_n(\vec{x}) \zeta_n^{\dagger}(y^-,x_{\perp})}{2L\lambda_n} .
\end{equation}
The Green's function defined in the main text follows trivially
\begin{equation}
{\cal G}_{(-)}[\vec{x},\vec{y};V_-] = \delta^{(2)}(x_{\perp}-y_{\perp})
{\cal G} (x^- , y^-; x_\perp) .
\end{equation}
It should be emphasized that the basis functions $\zeta_n$ as well as
the eigenvalues $\lambda_n$ depend on the dynamical variable $V_-$.
The Green's function satisfies
\begin{equation}
iD_- {\cal G} (x^-, y^-) = \delta (x^- - y^-) - \frac{1}{2L} \sum_{n_0} 
\zeta_{n_0}(\vec{x}) \zeta_{n_0}^{\dagger}(y^-,x_{\perp}).
\end{equation}
Recall that $\lambda_{n_0}=0$. The simplest example is, of course,
$V_-=0$. Then $n_0=0$ and the corresponding eigenfunction is
$\zeta_0=1$.
Thus (including the perpendicular delta function)
\begin{equation}
i \partial_- {\cal G}_{(-)}[{\vec x}, {\vec y};0]
= \delta^{(2)}(x^{\perp}-y^{\perp})
\left[ \delta(x^- - y^-) - \frac{1}{2L} \right]\;,
\end{equation}
and
\begin{equation}
{\cal G}_{(-)}[{\vec x}, {\vec y};0] = -i \delta^2(x_\perp-y_\perp)
 \sum_{n \ne 0} \frac{1}{2i\pi n}
\exp \left(\frac{2i\pi n (x^- - y^-)}{2L}\right) \;,
\end{equation}
where one readily recognises the periodic step function $\tilde{\vartheta}$
in the sum on the right hand side.
Analogous expressions can be found for antiperiodic functions
by replacing $2n \rightarrow (2 n + 1)$ in 
Eqs.(\ref{opereigeneq},\ref{eigenfun},\ref{eigenval}). For
the example above,  $V_-=0$, one obtains
\begin{equation}
i \partial_- {\cal G}_{(-)}^{a}[{\vec x}, {\vec y};0]
= \delta^{(2)}(x_{\perp}-y_{\perp})
\delta^{a}(x^- - y^-) \;,
\end{equation}
and
\begin{equation}
{\cal G}_{(-)}^{a}[{\vec x}, {\vec y};0] = -i \delta^2(x_\perp-y_\perp)
 \sum \frac{1}{(2n+1)i\pi}
\exp \left(\frac{(2n+1)i\pi (x^- - y^-)}{2L}\right) 
\end{equation}
for antiperiodic boundary conditions.

With the explicit form of the Green's function it is straightforward
to study the effect of the unitary gauge fixing transformations on it, 
thereby verifying the results given in the main text. 
Consider, for example,
the first unitary transformation. As it is, ${\cal G}_{(-)}$ is 
obviously invariant
under the $U_1$ of Eq.(\ref{firsttrans}). 
However, since it is `sandwitched' between the fermion
operators $\xi$, it picks up the (dynamical) phases $e\vartheta$. 
One obtains
\begin{eqnarray}
&\;&{\cal G}_{(-)}[\vec{x},\vec{y}; V_-] \rightarrow
\exp\left(ie\vartheta(\vec{x})\right)
{\cal G}_{(-)}[\vec{x},\vec{y}; V_-]  
\exp\left(-ie\vartheta(\vec{y})\right)
\nonumber \\
&=& 
\exp\left(-e\int d^3 z \;{\cal G}_{(-)}[\vec{x},\vec{z};0] 
A_-(\vec{z})\right)
{\cal G}_{(-)}[\vec{x},\vec{y}, V_-]  
\exp\left(-e\int d^3 z \;{\cal G}_{(-)}[\vec{x},\vec{z};0]
 A_-(\vec{z})\right)
\nonumber \\
&=& 
\exp\left(ie\int d z^- \;\tilde{\vartheta} (x^- - z^-) 
A_-(x_{\perp},z^-)\right)
{\cal G}_{(-)}[\vec{x},\vec{y}; V_-]  \times \nonumber\\
&& \hspace{6cm}
\exp\left(-ie\int d z^- \;\tilde{\vartheta} (y^- - z^-) 
A_-(y_{\perp},z^-)
\right) \nonumber\\
&=& {\cal G}_{(-)}[\vec{x},\vec{y}; a_- + q_-]  
\end{eqnarray}
in explicit detail. 

\setcounter{equation}{0}

\section{Appendix B. Decomposition of electromagnetic currents.}
The electromagnetic current
\begin{equation}
{\cal J}^\mu = - e \bar{\psi} \gamma^\mu \psi
\end{equation}
satisfies periodic boundary conditions and therefore can be
decomposed into its global zero modes 
\begin{eqnarray}
Q^+ & = &  -  e\sqrt{2} \int 
\frac{d^3x}{8L\ L^2_\perp}
\psi^{\dagger}_{+}(\vec{x})\psi_{+}(\vec{x})\\
Q^{-} & = & -  e\sqrt{2} \int 
\frac{d^3x}{8L\ L^2_\perp}
\psi^{\dagger}_{-}(\vec{x})\psi_{-}(\vec{x})\\ 
Q^i & = & -  e \int 
  \frac{d^3x}{8L\ L^2_\perp}
\left[\psi^{\dagger}_{+}(\vec{x})\alpha^{i}\psi_{-}(\vec{x}) +
\psi^{\dagger}_{-}(\vec{x})\alpha^{i}\psi_{+}(\vec{x})\right]\label{defqi},
\end{eqnarray}
proper zero modes
\begin{eqnarray}
j^{+}(x_\perp) & = & -  e\sqrt{2}
\int_{-L}^{L} \frac{dx^{-}}{2L}
\psi^{\dagger}_{+}(\vec{x})\psi_{+}(\vec{x}) - Q^+\\ 
j^{-}(x_\perp) & = & - e\sqrt{2}
\int_{-L}^{L} \frac{dx^{-}}{2L}
\psi^{\dagger}_{-}(\vec{x})\psi_{-}(\vec{x})- Q^{-}\\ 
j^{i}(x_\perp) & = & -  e \int_{-L}^{L} \frac{dx^{-}}{2L}
\left[ \psi^{\dagger}_{+}(\vec{x})\alpha^{i}\psi_{-}(\vec{x}) + 
\psi^{\dagger}_{-}(\vec{x})\alpha^{i}\psi_{+}(\vec{x})\right] - Q^i
\label{defji} \end{eqnarray}
and the normal modes 
\begin{eqnarray}
J^{+}(\vec{x}) & = & -  e\sqrt{2} \psi^{\dagger}_{+}(\vec{x})\psi_{+}(\vec{x})
- j^{+}(x_\perp) - Q^{+}\\
J^{-}(\vec{x}) & = & - e \sqrt{2} 
\psi^{\dagger}_{-}(\vec{x})\psi_{-}(\vec{x}) - j^{-}(x_\perp) - Q^{-}\\
J^{i}(\vec{x}) & = & - e \left[
\psi^{\dagger}_{+}(\vec{x})\alpha^{i}\psi_{-}(\vec{x}) + 
\psi^{\dagger}_{-}(\vec{x})\alpha^{i}\psi_{+}(\vec{x})\right] -
j^{i}(x_\perp) - Q^{i}\;.
\end{eqnarray}
Through the constraint for $\psi_-$ these will involve the photon
degrees of freedom.
 
\setcounter{equation}{0}

\section{Appendix C. Momenta, Constraints and Dirac
Brackets.} 
Using the fermionic Lagrangian Eq.(\ref{fermionlagr}) one has the
canonical momenta 
\begin{eqnarray}
\Pi_{\psi_{+}} & = &\displaystyle \frac{\delta L_{fer}}{\delta
(\partial_{+} \psi_{+})} = - i\ \sqrt{2}
{\psi_{+}}^{\dagger} \label{defpipsip}\\
\Pi_{\psi_{-}} &  = & \displaystyle \frac{\delta L_{fer}}{\delta
(\partial_{+} \psi_{-})}  =  0\\
\Pi_{{\psi_{-}}^{\dagger}} & = & \displaystyle \frac{\delta
L_{fer}}{\delta (\partial_{+} {\psi_{-}}^{\dagger})}  =  0\\
p^i & = & \displaystyle \frac{\delta
L_{fer}}{\delta (\partial_+ q_{i})} = 0.
\end{eqnarray}
The extended canonical Hamiltonian density defined as
\begin{eqnarray}
{\cal H}^{fer}_E = (\partial_{+}\psi_{+}) \Pi_{\psi_{+}} 
 - {\cal L}_{fer} + \Pi_{\psi_{-}} v +
{v}^{\dagger}\Pi_{{\psi_{-}}^{\dagger}} + 
{u_i} {p^i},
\end{eqnarray}
after some simple manipulations takes the form
\begin{eqnarray}
{\cal H}^{fer}_E & = &
-  i \sqrt{2} {\psi_{-}}^{\dagger} \partial_{-}
\psi_{-} - i {\psi_{-}}^{\dagger} \alpha^i \partial_{i}
\psi_{+}  - i {\psi_{+}}^{\dagger} \alpha^i \partial_{i}
\psi_{-}\nonumber\\
&+ & m {\psi_{+}}^{\dagger} \gamma^0 \psi_{-} + m
{\psi_{-}}^{\dagger} \gamma^0 \psi_{+}+  e \sqrt{2}
{\psi_{-}}^{\dagger} \psi_{-}V_{-} \label{extHamfer}\\
&+& e V_i \left( {\psi_{+}}^{\dagger} \alpha^i \psi_{-}+
{\psi_{-}}^{\dagger} \alpha^i \psi_{+}\right) + \frac 1 2
j^i {\cal G}_{(\perp)}[0]\ast j^i \nonumber\\
& + & \Pi_{\psi_{-}} v + {v}^{\dagger}\Pi_{{\psi_{-}}^{\dagger}} 
u_i p^i \nonumber
\;.
\end{eqnarray}
In this system, there are primary constraints
\begin{eqnarray}
\Phi_{1} & = & \Pi_{\psi_{-}} \\ 
\Phi_{2} & = & \Pi_{\psi_{-}^{\dagger}} \\
\Phi_{3}^i & = & {p^i}
\end{eqnarray}
and the secondary constraints, which are given by the stationarity
conditions 
\begin{eqnarray}
\chi_1 & = &\partial_{+} \Phi_{1}  =  i \sqrt{2} {\partial_{-}}
{\psi_{-}}^{\dagger} + i {\partial_i}{\psi_{+}}^{\dagger}
\alpha^i + e \sqrt{2} {\psi_{-}}^{\dagger} {V}_{-} \nonumber\\ 
&& + e {\psi_{+}}^{\dagger} \alpha^i \left( V_i - {\cal G}_{(\perp)} \ast
{j^{i}} \right) + m {\psi_{+}}^{\dagger} \gamma^0\simeq 0 
\label{chi1}\\ 
\chi_2 & = & \partial_{+} \Phi_{2} = i \sqrt{2} \partial_{-}
\psi_{-} + i \alpha^i \partial_i \psi_{+} - e \sqrt{2}
\psi_{-} V_{-} \nonumber\\ 
&& - e \alpha^i \left( V_i - {\cal G}_{(\perp)} \ast
{j^{i}} \right) \psi_{+} - m \gamma^0
\psi_{+} \simeq 0\\ 
\chi_3^i & = & \partial_{+} \Phi_{3}^i = Q^i = - e \int d^3{x}
\left( \psi_{-}^{\dagger} \alpha^i \psi_{+} + \psi_{+}^{\dagger}
\alpha^i \psi_{-} \right)(\vec{x}). \label{chi3}
\end{eqnarray}
They all form a set of second-class constraints and Dirac brackets can
be given for nonconstrained variables
\begin{eqnarray}
\left \{ \psi_{+}^{\dagger}(\vec{x}),
{\psi_{+}}^{\dagger}(\vec{y}) \right \}^{+}_D & = & -
\frac{i}{\sqrt{2}} \ \Lambda^{(+)} \delta^{(3)}_a(\vec{x}- \vec{y})
\end{eqnarray}
while all others vanish. 

For the total effective lagrangian Eq.(\ref{effQEDLagr}) one finds
the canonical momenta 
\begin{eqnarray}
\Pi^{-} & = &\displaystyle \frac{\delta
L_{eff}}{\delta (\partial_{+} A_{-})} =
\Pi \\ 
\Pi^{i} & = & \displaystyle \frac{\delta
L_{eff}}{\delta (\partial_{+} A_{i})} =
\partial_{-} A_{i}\\ 
\Pi_{\psi_{(+)}} & = &\displaystyle \frac{\delta L_{eff}}{\delta
(\partial_{+} \psi_{(+)})} = - i\ \sqrt{2}
{\psi_{+}}^{\dagger} \\
\pi^{-}(x_\perp) & = & \displaystyle \frac{\delta L_{eff}}{\delta
(\partial_{+} a_{-}(x_\perp))} =  \pi(x_\perp). \label{defpipm}\\
p^{-} & = & \displaystyle \frac{\delta
L_{eff}}{\delta (\partial_{+} {q}_{-})} = 8 L L_\perp^2 
 \partial_{+} {q}_{-}.
\end{eqnarray}
There is only one primary constraint here, but it is 
second-class and renders only one Dirac bracket
(\ref{AppAiAipb}) different
from the corresponding Poisson bracket
\begin{eqnarray}
\left \{{\Pi}^{-}(\vec{x}),
{A}_{-}(\vec{y}) \right \}_D & = & - \delta_n^{(3)}
(\vec{x} - \vec{y})\label{Pbapi2}\\ 
\left \{ \psi_{(+)}^{\dagger}(\vec{x}),
{\psi_{(+)}}(\vec{y}) \right \}^{+}_D & = & -
\frac{i}{\sqrt{2}} \ \Lambda^{(+)} \delta^{(3)}_a(\vec{x}- \vec{y})\\
\left \{ \pi(x_\perp), {a}_{-} (y_\perp) \right \}_D & = & - 
\delta_p^{(2)}(x_\perp - y_\perp)\\ 
\left \{{A}_{i}(\vec{x}), {A}_j (\vec{y}) \right \}_D &
= & - \frac i 2 \delta_{ij} {\cal G}_{(-)}[\vec{x}, \vec{y}; 0]\label{AppAiAipb}\\ 
\left\{ p^{-}, 
{q}_{-} \right \}_D & = & - 1
\;.
\end{eqnarray}
With these, the final quantum commutators can be formulated, as given
in the text.

\setcounter{equation}{0}

\section{Appendix D. Gau{\ss} Law Operator, Residual Gauge Freedom}

In the Weyl gauge, the Gau{\ss} law operator $G(\vec{x})$
\begin{equation}
G(\vec{x}) = \partial_{-} \Pi(\vec{x}) + 2 \partial_{-} \partial_i
{A}_i (\vec{x}) - \Delta_\perp {A}_{-}(\vec{x}) -
\Delta_\perp {a}_{-} (x_\perp) - e \sqrt{2}
\psi^{\dagger}_{+}(\vec{x}) \psi_{+}(\vec{x}) 
\end{equation}
is $x^{+}$-independent by the Hamilton equations
of motion, i.e. it commutes with the Hamiltonian.
The independent fields have the following
commutators with $G(\vec{x})$
\begin{eqnarray}
\left [G(\vec{x}), {A}_{-}(\vec{y}) \right ] & = & - i
\partial^x_{-} {\delta_n^{(3)}} (\vec{x} - \vec{y})\\
\left [G(\vec{x}), {A}_{i}(\vec{y}) \right ] & = & - i
\partial^x_{i} {\delta_n^{(3)}} (\vec{x} - \vec{y})\\
\left [G(\vec{x}), \Pi(\vec{y}) \right ] & = &  - i
\Delta_\perp {\delta_n^{(3)}} (\vec{x} - \vec{y})\\
\left [G(\vec{x}), \psi_{(+)}(\vec{y}) \right ] & = & e 
\psi_{+}(\vec{x}) {\delta_a^{(3)}} (\vec{x} - \vec{y})\\
\left [G(\vec{x}), {a}_{-}(y_\perp) \right ] & = & 0\\
\left [G(\vec{x}), \pi(y_\perp) \right ] &
= &  - i \Delta_\perp\delta_p^{(2)} (x_\perp - y_\perp)\\
\left [G(\vec{x}), {a}_{-}(y_\perp)
\right ] & = & 0\\
\left [G(\vec{x}), p^{-} \right ]
& = & \left [G(\vec{x}), q_{-} \right ] = 0.
\end{eqnarray}	
The Gauss law operator is the generator of the residual
$x^{+}$-independent gauge transformations with periodic gauge
function $h(\vec{x})$: 
\begin{eqnarray}
{A}_i(\vec{x})^h & = & \Omega_h {A}_i(\vec{x})
\Omega^{\dagger }_h = {A}_i(\vec{x})  - \partial_i {h}(\vec{x})\\
{A}_{-}(\vec{x})^h & = & \Omega_h {A}_{-}(\vec{x})
\Omega^{\dagger }_h = {A}_i(\vec{x})  - \partial_{-} h(\vec{x})\\
\Pi(\vec{x})^h & = & \Omega_h \Pi(\vec{x}) \Omega^{\dagger }_h
= \Pi(\vec{x})  + \Delta_\perp h(\vec{x})\\
\psi_+(\vec{x})^h & = & \Omega_h \psi_+(\vec{x}) \Omega^{\dagger }_h
= e^{ i e h({\vec x})} \psi_+ (\vec{x})\\
{a}_{-}(x_\perp)^h & = & \Omega_h {a}_{-}(x_\perp)
\Omega^{\dagger }_h = {a}_{-}(x_\perp)  \\
\pi(x_\perp)^h & = & \Omega_h \pi(x_\perp)
\Omega^{\dagger }_h = \pi(x_\perp) + \Delta_\perp
\int_{-L}^{L} dy^{-}\ h(y^{-}, x_\perp)\\
{p^{-}}^h & = & \Omega_h p^{-}
\Omega^{\dagger }_h = p^{-}  
\end{eqnarray}
where
\begin{equation}
\Omega_h = e^{i \int d^3{\vec x} h({\vec x}) \ G(\vec{x})}.
\end{equation}
In the above transformations for fermion fields the following
identity has been used
\begin{equation}
\psi_{+}(x^{-}){\delta}_a(x^{-} - y^{-}) =
\psi_{+}(y^{-})\left[ {\delta} (x^{-} - y^{-}) +
\frac{1}{2L}\right]
\end{equation}
where the delta distribution on the right hand side
is now the complete expression for periodic functions.
One can prove this by
integrating its both sides either with a smooth test function
periodic in the $x^{-}$ variable or a smooth test function
antiperiodic in the $y^{-}$ variable.

 \end   {document}